\documentclass[iop,apj]{emulateapj}
\usepackage{amsmath,amssymb,amstext}

\usepackage[breaklinks,colorlinks,citecolor=blue,linkcolor=magenta]{hyperref}
\usepackage[all]{hypcap} 

\usepackage{multirow}
\shorttitle{Merger-driven direct collapse}
\shortauthors{Mayer \& Bonoli}

\begin{document}

\title{The route to massive black hole formation via merger-driven  direct collapse: a review}

\author{Lucio Mayer}
\affiliation{Center for Theoretical Astrophysics and Cosmology, Institute for Computational 
Science, University of Zurich,
Winterthurerstrasse 190, CH-8057 Zurich, Switzerland, and Kavli Institute for Theoretical 
Physics, Kohn Hall, University of California, Santa Barbara, CA 93106, USA, lmayer@physik.uzh.ch}

\author{Silvia Bonoli}
\affiliation{Centro de Estudios de Fisica del Cosmos de Aragon, Plaza San Juan 1, Planta-2, 
E-44001, Teruel, Spain}

\begin{abstract}

The direct collapse model for the formation of massive black holes 
has gained increased support as it provides a natural explanation
for the appearance of bright quasars already less than a billion years from
the Big Bang. In this paper we review a recent scenario for direct collapse that
relies on multi-scale gas inflows initiated by the major merger of massive gas-rich
galaxies at $z > 6$, where gas has already achieved solar composition.  
Hydrodynamical simulations undertaken to explore our scenario show
that supermassive, gravitationally bound compact gaseous  disks weighing a billion solar masses,
only a few pc in size, form in the nuclei of merger remnants in less than $10^5$ yr.
These could later produce a supermassive protostar or supermassive
star at their center via various mechanisms. 
Moreover, we present a new analytical model,
based on angular momentum transport in mass-loaded gravitoturbulent disks. 
This naturally predicts that a nuclear disk accreting at rates
exceeding $1000 M_{\odot}$/yr, as seen in the simulations,
is stable against fragmentation irrespective of its metallicity.
This is at variance with conventional direct collapse
scenarios, which require the suppression of gas cooling in metal-free protogalaxies for gas collapse to take place.
Such high accretion rates reflect the high free-fall velocities in massive halos appearing only at $z < 10$, 
and occur naturally as a result of the efficient angular momentum loss provided by the merger dynamics.
We discuss the implications of our scenario on the
observed population of high-z quasars and on its evolution to lower redshifts using
a semi-analytical galaxy formation model. Finally, we consider the intriguing
possibility that the secondary gas inflows in the unstable disks might drive
gas to collapse into a supermassive black hole directly via the General Relativistic
radial instability. Such {\it dark collapse} route could
generate gravitational wave emission detectable via the future Laser Interferometer Space Antenna (LISA).
\end{abstract}

\maketitle

\section{Astrophysical Scenarios for massive BH formation and their and Challenges}

The first evidence on the existence of supermassive black holes, with masses in the range
$10^6-10^{10} M_{\odot}$, comes from the discovery of quasars and radio galaxies in the 60s
(Schmidt 1963).
 When X-ray satellites became available, many of these sources revealed also
powerful X-ray emission (see, e.g., the review of Brandt \& Hasinger 2005). Galaxies  presenting this kind of
activity are generically referred to as Active Galactic Nuclei (AGNs), as the
activity originates from the nuclear region of the galaxy itself.
It was soon realized that the most natural way to explain the gargantuan luminosities
associated with such astrophysical phenomena, which can outshine the galaxies that host them,
is the  energy radiation derived from the loss of potential energy of matter as it falls onto
a massive compact object located in the galactic nucleus (Salpeter 1964; Lynden-Bell 1969).
This {\it accretion} process, which
is driven purely by the gravitational pull of the compact nuclear object,  is known
to be the most efficient way to convert a mass reservoir into heat, and then radiation.
It is significantly more efficient than nuclear fusion in stars (Shapiro \& Teukolski 1987).
In the present-day Universe
we have evidence that massive black holes also exit in galaxies with no signs of
nuclear activity.
This evidence comes from the dynamical study of the motion of stars and gas in galactic nuclei
in the region where the central black hole dominate the gravitational potential
(the, so, called, {\it sphere of influence}\footnote{The black hole sphere
of influence is typically defined to be a sphere with a radius
$r_{h}=GM_{BH}/\sigma^2$, where $\sigma$ is the central velocity dispersion of
the host galaxies. For a typical black hole of $10^8 M_{\odot}$ in a galaxy of
$\sigma \sim 200 km/s $, the sphere of influence has a radius of $\sim 11 $pc.} of the black hole). For the
Galacic Center we have the most exquisite determination of the central black hole mass
from studying individual orbits of stars well within the gravitational influence
radius of the black hole (e.g., Gillessen et al. 2017).
 The combination of dynamical analysis for the study of ``dormant'' black holes,
and the study of the frequency of AGN, led to 
 the conclusion that massive black
holes are ubiquitous at the center of galaxies over a broad range
of masses (Kormendy \& Ho 2013), from the most massive early-type galaxies at the centers of
groups and clusters to the relatively low-mass galaxies the size of the Large Magellanic
Cloud (Pardo et al. 2016).

Many models have been proposed for the
origin of massive black holes (see the recent reviews by Volonteri 2012; Haiman 2013;
Johnson \& Haardt 2016, and
Latif \& Ferrara 2016), where the different formation scenarios postulate different
types of precursors (or ``seeds'') for massive black holes.
 After formation,
 seed black holes start experiencing phases of exponential growth, which
generally wash away
 direct information on the initial mass. 
 The initial mass of the seeds depends on the type of precursor
and on the processes associated to the formation of the black hole. 
The most straightforward,  and widely explored,
scenario assumes that the precursors of black hole seeds are Population III
stars, the first generation of stars, appearing in the first hundred
million years of the history of the Universe (e.g., Abel et a. 2002; Bromm \& Larson 2004;
Greif et al 2010; Hirano et al. 2014).
Despite the large masses reached by PopIII stars ($100-1000
M_{\odot}$), the seed black holes resulting by the collapse of such stars would be orders of magnitude
below the mass of supermassive black holes (Madau \& Rees 2001).
 Because of the nature of the precursor and the relatively small mass of the
initial black hole, this scenario for the origin of
massive black holes is usually referred to as  ``PopIII'' or ``light seeds'' 
scenario.
Alternative theories have explored models in which seeds form from the collapse
of very large
gas clouds, with a significantly larger initial mass ($> 10^4 M_{\odot}$).
Depending on the type of evolution of the precursor gas cloud, the upper limit
for the seed mass could range from $10^6 M_{\odot}$ to $10^9 M_{\odot}$ in some
 very extreme scenarios (e.g., Mazzucchelli et al. 2017).
These models, commonly referred to as ``direct collapse'' or  ``heavy seed''
models, have gained more popularity in the recent years, as more evidence is
building up on the existence of extremely massive black holes at very high redshifts,
and such heavy seeds provide a significant ``head start'' for the mass build-up of
the black holes powering the most luminous quasars. 
Indeed, one important constraint on formation scenarios
comes from the existence of very bright quasars, with bolometric luminosities in the range
$10^{47}-10^{48}$ erg/s, which are already present at redshifts as high az $z \sim 7.1$ (Fan et al. 2006;
Mortlock e al. 2011;  Ba\~nados et al. 2018).  Even assuming that these quasars 
emit near the Eddington limit\footnote{ The Eddington
limit expresses the balance between radiation pressure force and the inward
gravitational pull exerted by the black hole. The Eddington luminosity, reached
at the limit, is given by $L_{\it Edd} = 4 \pi G M_{\it BH} \mu m_{p} c/
\sigma_{T}$, where $M_{\it BH}$ is the black hole mass, $\mu$ is the mean
molecular weight of the gas, $m_{p}$ is the mass of the proton and $\sigma_{T}$ is the
Thomson scattering cross-section.  For pure ionized hydrogen, the Eddington luminosity
is: $L_{\it Edd} = 1.28 \times 10^38 (M/M_{\odot})$ ergs/s.}, the masses inferred for the black holes are
in excess of $10^9 M_{\odot}$,
namely comparable to the most massive black holes in the present-day Universe.
The first problem that arises then is how to explain that such large masses can be achieved
so early on. At $z=7.1$ only 0.5 Gyr have elapsed since the Big Bang. A simple calculation
shows that one needs a seed black hole of at least $10^4 M_{\odot}$ in order to be able
to grow to such high masses while not overshooting the Eddington limit (see the
discussion in Section \ref{sec:timescales}).

Finally,  another set of scenarios, which can lead to seeds with masses in between the Pop III and the
direct collapse models, are those that involve core collapse of a dense star cluster (e.g., Portegies
Zwart et al. 2004; Devecchi et al. 2012).
Under this framework, the seed arises from the runaway mergers of stellar mass black holes and/or the formation of a central
Very Massive Star (VMS) from stellar mergers. Seeds with masses up to $1000 M_{\odot}$
can be formed in this scenario (see the review by Latif \& Ferrara 2016).

In the reminder of this section we will spend some time recalling the modern view
on galactic structure and discussing quantitatively
the important arguments concerning the accretion timescale of massive black holes
as well as the general features of the Pop III seed scenario and of direct
collapse scenarios. They will be instrumental to understand the astrophysical
motivations behind our new direct collapse scenario.

\subsection{A few remarks on galactic structure, mass transport and scaling
relations}

Our modern view of galaxies is that they comprise stars (distributed in
different morphologies), interstellar gas, a dark halo, which is dominant in
mass, and, in most cases, a central massive black hole. Our own
Milky Way conforms with this general description.
While  star formation is a distributed process, namely
stars form throughout the galaxy disk, massive black holes seem to be able to
assemble only in galactic
nuclei. Galactic nuclei contain typically less than $0.1 \%$ of the galaxy stellar mass in present-day galaxies
even in galaxies containing the densest and more concentrated stellar components, namely those hosting
a nuclear star cluster  (eg., Walcher et al. 2005; Georgiev et al. 2016). Nearby examples are
the nuclear star clusters in the late-type spiral galaxies  M33  and NGC 4244 (Seth et al. 2008). 
Instead SMBHs often comprise a few percent (up to almost 10\% in some cases) of the galaxy stellar mass
(Kormendy \& Ho 2013). Above the mass scale of M31 galaxies that have both a nuclear star cluster and a central
SMBH do indeed show that the latter has a mass up to 1000 times larger than the former (Figure 7 of Georgiev
et al. 2016).
In order for central massive holes to capture a much higher baryonic mass
relative to nuclear star clusters, one obvious source of their efficient growth must be the interstellar gas,
as this contributes at least a few percent of the stellar mass budget.
The latter however, is mostly found outside the nuclear region.
This leads to the notion of gas inflow, whereby baryonic matter at large radii
is transferred to the nuclear region as long as it is able to lose angular momentum
and sink. If star formation is negligible, such gas could  all add up to the mass
of the black hole. From this point of view, star formation and BH seed formation and/or  growth can be viewed  
as competing processes. 

We note that  significant effort in the community has been devoted to devise
scenarios in which star formation is heavily suppressed, thus  allowing a larger fraction of baryons
to turn into the central black holes seeds at early times. At the same time, as gas in galactic disks
is prevented to fall inwards by rotational support, generating inflows to the center requires to
find mechanisms to generate angular momentum transport (important also in the
context of black hole feeding,  see e.g., Jogee 2006). This is a very generic problem in many areas
where astrophysical disks are relevant, the prototypical example being accretion disks of stars and compact
objects, black holes included (e.g., Shapiro \& Teukolski 1987).

The body of theoretical knowledge developed in the context of star
formation and accretion disk physics will become relevant also for our purpose.

Relative to the mass budget of other components, at high redshift the conversion efficiency of baryonic
matter might have been even more skewed in the direction of  central black hole assembly, at least
in the galactic hosts of high-z quasars. Indeed, although
often not emphasized, the mass inferred for the black holes powering high-z quasars (up
to $10^{9}-10^{10} M_{\odot}$), could be comparable to that
of their host galaxies at that time, assuming that these galaxies are in the
mass range of what can be  inferred by abundance
matching  (e.g. Behroozi et al. 2013) given the halo masses ($10^{12}-10^{13}
M_{\odot}$) derived by
quasar clustering studies (see, e.g., the recent results at $z>3$ of Timlin et al. 2017 and He
et al. 2017). 
Given those numbers, it follows that at least 1\% of the standard
baryon budget of such halos is somehow converted into a massive black hole. The standard
baryon budget here corresponds to assuming that halos capture their share of baryons in proportion
to the cosmic baryon fraction suggested by different cosmological experiments
(e.g. Planck results,  Ade et al. 2015). This translates into
each halo having captured a baryonic mass of about $1/6$ of its dark matter mass. Now, for halos
of this mass various diagnostics, from abundance matching to weak lensing and clustering
analysis, suggest that, at least up to $z \sim 5$, the stellar mass of the galaxies embedded in them is also in the range
$2-6 \%$ of the halo mass (Behroozi et al. 2013; Moster et al. 2017). If these results can
be extrapolated to $z > 5$, which still awaits verification, it would mean that at early
epochs the black hole
formation efficiency could be as a high as the efficiency of the galaxy formation process
itself.  This is
quite striking since at low redshifts galaxies of similar halo and stellar masses, such as the Milky Way, have
black holes weighting only $10^{-6}$ of their halo mass. This seem to suggest that  the processes
at the origin of the high-z quasars could be  different from those responsible 
for the main population of massive black holes, and perhaps
quite exceptional. This could also imply
that the correlations between black hole masses and galaxy properties, such as stellar mass
or stellar velocity dispersion, widely
observed in low redshift galaxies, may not yet be established at $z > 5$ for the hosts of
the bright quasars. While evidence of the latter is
just beginning to be gathered, due to the obvious observational challenges, it is already clear that there can be substantial deviations from
the standard correlations already at $z \sim 4$ (e.g., Trakhtenbrot et al. 2015).

\subsection{The assembly  timescale of high-z quasars} \label{sec:timescales}

We briefly recall here the arguments that support the need for a massive BH seed in order
to explain the rapid emergence of the bright high-z quasars.
The time $t(M_{\it BH})$ required for a black hole of initial mass $m_{seed}$ to reach a mass
$M(BH)$, assuming Eddington-limited and continous accretion, is given by:
\begin{equation} \label{eqn:tBH}
 t(M_{\it BH}) = \frac{t_{\it Edd}}{f_{edd}} \frac{\epsilon_{r}}{(1-\eta)}
\ln \left( \frac{M_{BH}}{m_{seed}}\right),
\end{equation}
where $t_{\it Edd} \equiv c \sigma_{T}/(4 \pi G \mu m_{p})$ is the Eddinton time
($\sim 0.45$ Gyr for a pure hydrogen gas, with $\mu=1$),   $f_{edd}$ expresses 
at which fraction of the Eddington limit the black hole is accreting, and
$\epsilon_{r}$ and $\eta$ are, respectively,
the radiative efficiency and accretion efficiency\footnote{The accretion
efficiency $\eta$ directly depends on the spin of the black hole, and reaches
the maximum value of $\eta \sim 0.42$ for maximally rotating Kerr black holes.
The radiative efficiency $\epsilon_r$  depends both on the type of accretion
and on the accretion efficiency: for radiatively efficient accretion events, one
can assume $\epsilon_r = \eta$.}. Assuming a radiatively-efficient accretion
event, we assume $\epsilon_r = \eta = 0.1$, a typical average value for Shakura-Sunyaev
thin disks (Shakura \& Sunyaev 1973).
Adopting a more realistic molecular weight per electron for a plasma at zero metallicity with
cosmic abundance of hydrogen (X = 0.75) and helium (Y = 0.25), $\mu =
1/(1  - Y/2)
= 1.14$, the Eddington time lowers to $t_{\it Edd} \sim 0.39$ Gyr. Note that the
precise value of the metallicity is  marginal in this calculation, as
metallicity has only  a
 minor effect
on the value of the molecular weight (e.g., for a gas at solar metallicity,
$\mu=1.18$ instead of the value $\mu=1.14$ assumed here).
 The mass of the supermassive black holes powering the most luminous high-z
quasars is of the order of $M_{BH} = 10^9 M_{\odot}$, so that this is the value we
assume 
for the target mass.  Further assuming that the seed black hole is able to
accrete continuously at the Eddington limit  ($f_{edd} = 1$), the timescale
necessary to reach the target mass is 
$t_{(M_{BH})} = 0.4$ Gyr for  
 a seed of $10^5 M_{\odot}$, while is   $t_{(M_{BH})} \sim 0.7$ Gyr for a black
holes  starting from $10^2 M_{\odot}$, the  average value for light
seeds produced by Pop III stars. In this latter case the growth timescale is
comparable to the age of the universe at $z = 7$ (assuming the Planck cosmology
of Ade et al. 2015).
Hence, the rapid appearance of quasars within the first billion
years of the Universe can be more easily explained if the progenitors of these
luminous first quasars are massive seeds, and that is why the community has
devoted progressively more interest in models for the formation of massive
seeds, as further pointed out in the rest of this introductory section.

\subsection{PopIII seeds}

The first generation of stars, the primordial {\it Population III} stars, formed
in the first few hundreds
million years after the Big Bang. They are expected to have been  massive as the gas from which they
formed was metal-free, thus unable, via the efficient metal cooling, to cool to
low temperatures and fragment to the smaller
scales typical of ``normal'' stars (e.g., Abel et a. 2000; Bromm \& Larson 2004;
Greif et al. 2010). Indeed, in  metal-free
gas the only elements available are hydrogen and helium, which have limited radiative transitions compared to higher atomic number
elements.

Therefore, both Pop III remnants seeds and seeds formed in star clusters, with
masses typically below $1000 M_{\odot}$, might be
generally too light to grow to the most
massive black holes observed at high-z in the available time.
If the Eddington limit is superseded, then the accretion flow should be stopped and an outflow should
arise instead, as radiation pressure wins. Yet, this is true only in the simple spherical
isotropic accretion context, as we will discuss further in Section
\ref{sec:conclusions} . The environments where Pop III BH seeds may form
at $z \sim  15-30$ have been extensively studied with numerical simulations, which
agree on the fact that such seeds grow too slow, at Sub-Eddington rates, as the gas in their
host mini-halo has been efficiently photoionized and/or expelled by the massive stellar progenitor
(eg., Johnson \& Broom 2007; Wise et al. 2008).  This is because the primordial stars in the mass range $140-260
M_{\odot}$ explode as pair-instability supernovae (Heger et al. 2003),
liberating enough energy to effectively empty the mini-halos
of gas. On the other side, the lower-mass PopIII would still damp significant energy, having a significant
impact on the ambient medium if they form in clusters as it is likely (Greif 2015).
Only later, when star formation has led
to metal enrichment, increasing the cooling rate, and/or the original mini-halo has grown
significantly in mass to allow atomic cooling via Ly-alpha radiation, accretion can in principle
restart. If that happens, and Super-Eddington accretion can take place, fast growth might still
be possible on much less than a Gyr timescales (Lupi et al. 2016). Inayoshi, Haiman and Ostriker
(2016) found Super-Eddington flow  steady-state solutions provided that the density of the
surrounding gas is above a threshold density in order to trap photons. In this
case, a black hole
can grow continuously at very high rates even if accretion occurs spherically.
Still, no direct radiation hydrodynamical calculation
on nuclear or sub-nuclear  scales exists that shows that this growth mode can be
sustained up to the extremely  large BH masses  needed to  explain the bright
high-z quasars (see, e.g., the review by Johnson \& Haardt 2015). Therefore, the
existence of high-z quasars so far argues in favour of some form of direct collapse.
We note, however, that a broader, and less extreme, population of black holes
might have evolved from PopIII remnants, as further discussed in Section
\ref{sec:conclusions}. Two diverse channels of black hole formation could also
explain the dearth of lower-luminosity quasars seen by Treister et al. 2013.

\subsection{The different flavours of direct collapse} \label{sec:into_dc}

As mentioned above, with the term ``direct collapse'' the community generally refers to
all the different models that predict the formation of a ``heavy'' seed, with
 initial mass above the approximate value of $10^4 M_{\odot}$.  Ironically, multiple stages are involved
in the {\it direct} collapse scenarios, where the word {\it direct} refers to the
notion that there is no preceding phase of mass growth starting of a lighter initial
black hole. 
In this review we will focus on a particular flavour of direct collapse
which is different from the mainstream one. But before entering into  the
details of this new model
in the next section, we will  first describe here the more popular version of
direct collapse, as it will then be easier to appreciate the differences that our new model
introduces.

\subsubsection{Direct collapse in metal-free protogalaxies}

 In the standard picture, the key condition for the multi-scale gas
inflow necessary to generate
a central collapse into a massive black hole seed, is that fragmentation by gravitational
instabilities should be suppressed. The rationale behind requiring such condition is that
fragmentation would result into dense clumps that would produce regular stars, thus using up the
gas that would have otherwise ended up in the protogalactic nucleus.
 In this picture, the formation of the black hole seed has to precede star
formation. The necessity to avoid gas fragmentation translates into a condition on the
cooling time of the gas. Indeed, it is a general result of gravitational instability theory
in differentially rotating gas disks that fragmentation can only occur if local mass surface density
is high enough, so that gravity wins over pressure and centrifugal forces, and if the local
cooling time in such high surface density regions is shorter than the local orbital time
(Gammie 2001; Lodato \& Natarajan 2006). While for some time it has been difficult to quantify
exactly how strict such condition is in realistic three-dimensional disks, recent advances
in numerical techniques have shown that the condition is indeed really strict once numerical
sources of angular momentum loss are minimized (Deng, Mayer \& Meru 2017). In primordial gas
conditions, radiative cooling is provided mostly by molecular hydrogen ($H_2$)
as the metallicity of the gas should
be negligible, and thus the efficient cooling by fine structure lines of elements such as
carbon and oxygen, so important in the local interstellar medium at the present time, is
not relevant.  This is why a major line of investigation has been, and still is, to devise
ways by which molecular cooling can be stifled in protogalaxies. If this happens, it has
been shown by multiple studies that collapse can take place nearly isothermally, at a temperature
of the order of $10^4$ K (Regan et al. 2008; Choi et al. 2013;2015; Latif e al. 2013; Latif et al. 2015). 
The complete dissociation of molecular hydrogen requires the Lyman-Werner (LW) UV flux, namely photons
in the 11.2-13.6 eV energy band,  to be above
a certain threshold value, $J_{21}$, which depends on the radiation
spectrum (it is lower for hotter
radiation emitted by hotter, more massive stars) (Dijkstra et al. 2006; 2014; Agarwal et al. 2012; Latif et al.
2014; Visbal et al. 2014). Over the years, many studies aimed at quantifying
the threshold flux have clarified that the cosmic ionizing background at $z > 10$ is too low to
meet the requirements. $H_2$ dissociation could thus be possible only in the
presence of localized powerful ionizing sources.  
An option could be and internal ionizing source, but that requires on-going star formation,
which is what one wants to avoid in the first place in order to keep the gas metal-free.
The more promising alternative are external localized sources, such as nearby star forming
protogalaxy companions.
The proximity and duration of exposure of
the target galaxy to the luminous emitting nearby source (10-100 Myr) are crucial to achieve the desired
dissociation, although recent works estimate that galaxy pairs with the required configuration
should happen at least as frequently as the (rare) high-z quasars (Regan et al. 2017, but
see Habouzit et al. 2016 for a possible tension between required number densities and probability of
the required strength of the LW illuminating flux). Nevertheless,
in none of the  existing simulations, star formation in the galaxy source has been modelled
in a fully self-consistent way nor the role of gas metallicity has been properly addressed.
Latif et al. (2016) have shown that even with metallicities as small as
$10^{-4}$ solar, inflow
rates diminish drastically due to fragmentation and star formation if dust cooling is included.
 It is also still an open question
whether or not, with a nearby star forming galaxy, the target galaxy can remain completely
metal-free. Supernovae-driven outflows, which will frequently result as many of the massive stars
producing the LW flux die out,  can pollute the intergalactic medium well beyond
the host halo virial radius, and indeed this is considered the main driver of
metal pollution of the Intergalactic Medium (e.g., Dave et al. 2001).
Furthermore, the complete absence of fragmentation, even in metal-free,
dissociated media,
could not be a realistic outcome, as recent simulations show that the nuclei of protogalaxies are essentially rapidly
rotating disks that, provided that the numerical model has enough resolution, can fragment into binaries
or multiple objects at pc scales and below (Latif et al. 2015b). Interestingly, it has been argued that such
clumps might merge quickly at the center of the system before regular star formation ensues, so that
ultimately the same accumulation of gas can occur as when fragmentation is not present (Inayoshi \& Haiman 2014; Mayer et al. 2015).

Another potential issue with this conventional direct collapse picture is
related to the rate of gas inflow.  The estimated inflow rates 
 are of the  order of  $0.1 M_{\odot}$/yr, peaking at $1 M_{\odot}$/yr in some
cases (e.g., Latif et al. 2015). While it is unclear how such rates vary on long timescales
due to the limited time-baseline of the simulations, they are on the low-side compared
to those required by models in which black hole seeds form from the collapse of protostars
that avoid nuclear fusion, hence do not become supermassive stars, but rather
{\it quasi-stars}
(Begelman et al. 2006; Volonteri \& Begelman 2010), which allow fast growth of the initial
black hole seed.
Quasi-stars are radiation-pressure supported envelopes
in quasi-equilibrium with a central seed BH, essentially akin to giant stars in
which the force
supporting them against collapse is provided by accretion rather than fusion (Begelman et al. 2006;
Begelman 2010). We defer the reader to section \ref{sec:pathways} for a detailed discussion of the various
pathways to BH seed formation in direct collapse scenarios after a massive precursor has been
assembled. Here we recall that very high inflow rates are favoured
in certain  pathways to generate massive BHs, such as those
in which the radial General Relativistic instabilities directly bring a supermassive cloud or protostar to
dynamically collapse into a BH of the order of its mass, namely without prior formation of a
supermassive star (Fowler \& Hoyle 1964; Shibata \& Shapiro 2002; Ferrara et al.
2014).

The reason why the estimated inflow rates in the traditional direct collapse
scenario  are
relatively low, is due to the fact that this model of direct collapse has to
take place at high-z ($z>15-20$), so that the requirement of primordial gas
metallicity can be satisfied. At such redshifts, the possible host for direct
collapse would be the  atomic-cooling halos,  in the range $10^8-10^{9} M_{\odot}$. 
As explained in the next section, simple scaling arguments can also be used to
recover the relatively-low
inflow rates found in the simulations for halos in this mass range.

\subsubsection{The new direct collapse model}

A few years ago we began considering a whole different scenario for the
emergence of direct collapse seeds, in which no particular 
restrictions on the thermodynamical conditions of the gas phase are required.
Instead, the model relies on a particular trigger
of multi-scale inflows: galaxy mergers (Mayer et al. 2010).
In this alternative  picture, which  is the focus of 
this article, the event generating the conditions for BH seed formation in the galactic nucleus
is the major merger between two massive gas-rich galaxies.
Star forming galaxies  with stellar masses comparable to the ones of  massive
spirals, such as our own Milky Way at the present day, already exist at $z\sim 10$. 
Their host halo masses are correspondingly 3-4 orders of magnitude higher than the halo masses typically assumed
in simulations of metal-free protogalaxies, being in the range $10^{11} - 10^{12} M_{\odot}$ as opposed
to $10^{8} - 10^{9} M_{\odot}$.
These massive galaxies evolve in highly biased regions where the metal pollution by previous generation
of stars has proceeded already, hence the interstellar medium has a composition akin to normal present-day
galaxies, which implies efficient radiative cooling via metal lines (Mayer et al 2015).
This is a major difference relative
to the standard direct collapse route. Measurements of the gas metallicity in the hosts of some of the
bright high-z quasars, via detection of the CO molecule, do exist (Walter et al. 2004). They imply, indeed,
gas with already solar metallicity. Note that highly biased regions, corresponding to density fluctuations
whose amplitude is significantly higher than the mean amplitude (4-5 $\sigma$
peaks in technical jargon\footnote{In a hierarchical structure formation
scenario, structures forms from small density fluctuations that grow as the
Universe evolves. The amplitude of the fluctuations is defined as $\nu \equiv
\delta_c/\sigma(M,z)$, where $\delta_c \sim 1.69$ and $\sigma(M,z)$ is the rms
variance of the linear density field smoothed on a scale $R(M)=(3M/4\pi \rho_m)^(1/3)$ (Press \& Schechter 1974).}) are also implying 
the high clustering amplitude of the high-z quasars. Cosmological simulations that assume pre-existing massive BH seeds
in the range $10^4-10^5 M_{\odot}$ and concentrate on modelling their growth can do it successfully in
such highly biased regions as accretion of cold gas funnelled by cosmological filaments is very
high (e.g., Di Matteo et al. 2012; Feng et al. 2014; Di Matteo et al. 2017).
Furthermore, major mergers do occur
at $z \sim 8-12$ in these simulations as such massive early galaxies assemble
(Di Matteo et al. 2012), precisely because the  galaxy formation process is
accelerated in high sigma peaks (Wilkins et al. 2017).
In essence, our new scenario for the efficient accumulation of gas in the center
of high-z massive galaxies, connects naturally with the models of fast growth
of massive black holes, which successfully reproduce the rapid appearance of
bright quasars by $z \sim 7$, exploiting
the nature of the special environments in which quasars appear to exist at high redshift.
We finally underline that the high-mass halos involved in this model are also able to
sustain  gas inflow rates that are significantly higher than the ones possible
in $z\sim 15-20$ protogalaxies. This, as we will further see in the next section, has paramount consequences
on the nature of gas collapse at small scales, and thus on the nature of the precursor of the BH seed.
\\

The rest of the article is structures as follows. In the next two sections we review the details of our merger-driven model for direct collapse,
providing also a deeper look
 at the physics involved than in the specialized publications already published.
We start by elucidating important notions, such as the connection between gas infall rates and host halo
potential well and, most importantly, how the emergence of a dense, optically thick nuclear gas disk
stable to fragmentation might be the necessary consequence of heavy loading via high inflow rates. 
This will be the rationale behind our novel analytical model of massive loaded nuclear disks
presented in section 2. With these
premises it will be easier to understand the numerical results of our 3D hydrodynamical simulations
reported in section 3.  In section 4 we will outline the possible pathways 
for the evolution of  central massive central baryonic objects  towards a massive BH seed.
We will then connect our results to the large-scale cosmological context in
section 5, where  we present
a semi-analytical model of galaxy formation equipped with our direct collapse scenario to
study how well we can reproduce not only the high-z quasars but also the many other observables 
associated with the population of luminous AGNs as well as non-accreting massive BHs at both
high and low redshift. Finally, section 6 will contain a discussion of alternative rapid BH growth
models, focusing on Super-Eddington accretion, and will summarize the observational challenges in
constraining the various scenarios, and ours in particular.

\section{The physics of multi-scale gas inflows in high-redshift galaxies}.

In this section we review the physical processes related to the
formation of protogalactic disks during the first phases of structure formation,
and we will summarize the basic concepts related to the stability and evolution of these disks in
the case of no external perturbations. In Section 3 and 4 we will see how this picture
changes for the case of a nuclear disk formed during a major galaxy merger, in which
the infall of gas from galactic scales has a substantial effect on its dynamical evolution.
Section 4, in particular, will present a simple analytical framework that attempts to address
the physics behind the simulations described in section 3, giving rise to what we will dub ``the
loaded disk model''.

\subsection{Inflows at supra-galactic scales: from the dark halo to the protogalactic disk}

The first step that needs to be accomplished by any direct collapse scenario is
to efficiently bring interstellar gas from kiloparsec scales to parsecs scales and below. This is a necessary stage to
achieve the formation of a massive, ultra-dense baryonic core, a potential precursor of the BH seed.
The framework within which this has to be studied is the modern theory of galaxy
formation within  the
Cold Dark Matter (CDM) cosmological model (White \& Rees 1978). In CDM, cosmic structures form hierararchically,
in a bottom-up fashion, from gravitational instability of small initial density perturbations in 
a dissipationless dark matter component, which eventually grow into galaxy-sized dark halos by
via accretion of matter  and mergers with other halos (Blumenthal et al. 1980; White \& Frenk 1991). 
In CDM the predicted size scale of dark halos, the virial radius $R_{vir}$, embedding typical present-day galaxies such as our own Milky Way is in the range 200-300
kpc, and such halos have characteristic virial masses $M_{vir} \sim 10^{11}-10^{12} M_{\odot}$ (precisely, these are the typical numbers that
apply for a CDM model with a dominant contribution of the cosmological constant at low redshift, as it is
in the case of our Universe). 
At higher redshift, since the mean density
of the Universe is higher, halos  collapse at a higher characteristic density, resulting in a smaller $R_{vir}$,
which roughly scales as $1/({1+z})^{3/2}$. 
Baryonic matter 
falls into DM halos due to their gravitational attraction, cooling and forming a central rotating disk due to 
conservation of an initial small amount of angular momentum imparted by tidal
torques  (White \& Rees 1978). \\

The simplest approach to evaluate the importance of gas inflows in galaxy-scale systems is to
assume that gas is freely-falling inside the potential well of the host dark matter halo. 
In this case, one can borrow the basic concepts of the so called {\it inside-out collapse}
framework of star formation  theory (eg., Stahler \& Palla 2010). 
We thus envision that baryons first accrete in free-fall to the center of the halo
from the cosmic web, and, after a sizable mass has gathered, sustained accretion onto the
baryonic core continues due  to its own gravitational pull. 
Inside-out collapse refers to this second phase.
This simple model is reasonable in systems where (i) the  angular momentum is
negligible  and (ii) the cooling is efficient enough to yield nearly isothermal
conditions even in the presence of shocks or other
heating mechanisms during infall.
About the first requirement, spins of halos in CDM are small at the virial
radius scale, but eventually angular momentum has
a major impact on the gas flow at smaller scales, where a disk forms as fluid elements reach
the radius of centrifugal support (White \& Rees 1978). 
This  essentially means that the model will break down inside the protogalactic disk.
Note also that in CDM the spin of baryons is essentially imposed by the spin of the underlying dark
matter distribution, and has a well defined universal distribution (eg Bullock et al. 2001a) yielding
a centrifugal radius that is always 4-5\% of the virial radius $R_{vir}$. This implies that,
for galactic halos, the centrifugally supported disk that forms should have a scale of 10-15 kpc,
roughly consistent with present-day observed galaxy sizes, but would be much more compact in protogalaxies 
at high-z, $1-1.5$ kpc at z=8 for example, because of the reduction of $R_{vir}$ (Mo, Mao \& White. 1998).
The second condition, requiring the gas to be isothermal during  free-fall, 
can be satisfied in cosmological filamentary accretion from the cosmic
web, the so called {\it cold mode 
accretion},
which is believed to be the main mode of galaxy formation at least up  to halo virial masses of  
nearly $10^{12} M_{\odot}$ (see
Keres et al. 2009). Cold flow accretion is indeed the result of radiative cooling overtaking shock
heating to zero-th order (White \& Frenk 1991; Dekel \& Birnboim 2006) so that a stable
hot gaseous halo cannot form below the aforementioned mass threshold.
At very low mass scales photoionization by the cosmic ultraviolet background would
stop or at least slow down the infall, but this is relevant after reionization, namely below the
$z = 8-20$ redshift range that BH seed formation mechanisms usually consider.

At the boundary of the protogalactic disks we can determine the gas infall rate in the following
way:

\begin{equation}
\dot M = \alpha  {V_{ff}}^3/G
\label{eqn:Mdot}
\end{equation}

where the free-fall velocity is $V_{ff} \sim V_{circ}$, with $V_{circ}$ being the so-called halo circular 
velocity, namely the velocity
that a gas parcel would have on a circular orbit at a specific radius in the halo potential well (neglecting corrections
due to the additional mass of the baryons), i.e. $V_{circ}(R) =  \sqrt{(G M(R) / R)}$, where $G$
is the gravitational constant, and $M(R)$ the enclosed mass of the system at the
radius $R$. 
Finally, $\alpha$ parametrizes the viscosity of the flow (see 
Lodato \& Natarajan 2006).
We will also assume that $V_{circ} \sim V_{vir}$, where $V_{vir}$ is
the virial velocity of the dark halo, namely the characteristic velocity scale resulting by the
gravitational collapse of collisionless dark matter in an overdensity with a characteristic 
virial mass $M_{vir}$. This approximation neglects that the circular velocity is not
a constant, but first raises, and then decreases at large radii in CDM halos reaching
$V_{vir}$ at the halo boundary (Mo, Mao \& White 1998). Yet, such 
variations do not exceed a factor of 2-3 at the scales of interest, hence can be safely neglected.
As the overdensity amplitude sufficient for collapse depends on cosmological density
parameters, the above scaling equations will depend on cosmology in detail. When we provide specific numbers in this
review we will always assume the concordance cosmological model with matter density parameter $\Omega_M =  0.3$ and
cosmological constant density parameter $\Omega_{\Lambda} = 0.7$
Cosmological scaling laws between dark halo global parameters, such as virial
mass, virial radius and circular velocity, imply that  $V_{vir} \sim M_{vir}^{1/3}$ (Mo, Mao \& White 1998).
Moreover, we can assume $\alpha=1$ for the infalling matter described by eq.(1), as there is no natural source of 
bulk viscosity that would slow down the flow.
Instead, at smaller scales, if a gravoturbulent protogalactic disk is
established, a more appropriate value for the viscosity parameter is
$\alpha=0.1$ (see section 4 for a detailed justification of how
to choose the viscosity parameter).

Thus, assuming $c_s \sim V_{vir} \sim M_{vir}^{1/3}$ in eq. 3,
yields that $\dot M \propto M_{vir}$, highlighting 
that the inflow rate depends on the mass of the halo linearly. This has important implications as it suggests
that much larger inflow rate can be sustained  in higher mass halos. As in CDM theory cosmic
structures form hierarchically, in a bottom-up fashion, more massive halos form as time progresses, suggesting
that larger inflow rates are achieved at lower redshifts.
At $z \sim 15-20$, the typical redshift range considered in studies of first objects, the largest
halos that can cool via atomic cooling, will have masses between $10^9$ and $10^{10} M_{\odot}$, while
at $z \sim 8$ the first halos with virial masses of order $10^{12} M_{\odot}$ already appear, corresponding
to $4-5 \sigma$ density fluctuations above the typical amplitude (Di Matteo et al. 2012, Feng et al. 2016).
An increase of 2-3 orders of magnitude in halo masses means thus a corresponding increase of 2-3
orders of magnitude in inflow rate. Simulations of protogalaxies
at $z \sim 15-20$ show that inflow rates are of the order of $1 M_{\odot}$/yr  in $10^7 M_{\odot}$-$10^8 
M_{\odot}$
halos, hence at $z \sim 8$, if our scaling arguments apply, inflow rates could in principle increase to $> 1000 
M_{\odot}$/yr in $10^{12} M_{\odot}$ halos.
Once the gas joins the proto-disk, viscosity could reduce the inflow rate by an
order of magnitude (see also section 4), but in absolute terms the gas supply rate would still be remarkable.
This shows the potential advantage of considering lower redshift, more massive halos in triggering
prominent inflows.  Such prominent inflows could in principle assemble a sizable central cloud
by accumulating matter in the nucleus of the protogalaxy, which could then produce a massive BH
seed in various ways (see section 5). However, since CDM halos have
the same statistical distribution of spins irrespective of virial mass, the radius of 
centrifugal support is typically 3-4 \% of the virial radius $R_{vir}$ at all mass scales, with $R_{vir} \sim V_{vir}$
(Mo et al. 1998). As a result, in higher mass halos the centrifugal barrier occurs at a much larger scales,
which could potentially prevent the inflow from reaching pc scales and below. This is true unless some
mechanism causes a catastrophic loss of angular momentum of the baryons, essentially reducing the
centrifugal radius by orders of magnitude. As the simulations in section 3 demonstrate, this is what
the strong torques and hydrodynamical shocks in the merger between two nearly equal mass galaxies can do (Kazantzidis et al. 2005),
The conventional argument, though, in discarding lower redshift halos has always
been that these would be metal-enriched, undergo then fast cooling and fragmentation, which would
stifle inflows. Before going into the specific case of our merger model we will discuss, in the
next section, how this argument on fragmentation is misleading since it neglects the rich and  complex dynamics
of self-gravitating flows. Self-gravitating disks have been widely studied with 3D simulations in the case of star 
and planet formation (Durisen et al. 2007). We can borrow many useful results and knowledge from these other 
fields.

\subsection{Stability and fragmentation of self-gravitating disks}\label{sec:stability}

As we explained in the Section \ref{sec:into_dc}, the efficiency of radiative
cooling plays a key role in conventional direct
collapse models because  it is directly tied to the susceptibility to fragmentation.
 How much a rotating, self-gravitating fluid system is prone  to fragmentation, can be described by the combination 
of i) the Toomre Q parameter (Toomre 1964) parameter, which describes the
balance between the stabilizing and de-stabilizing forces acting on the fluid
and  ii)  considerations on the ability of the system to remain in
a stable or unstable state based on its radiative cooling rate (Gammie 2001).
The Toomre Q parameter is defined as $Q = \kappa c_s/ \pi G \Sigma$, and can be derived using the
framework of linear perturbation theory assuming a local axisymmetric
perturbation in a razor-thin
rotating axisymmetric disk (essentially a two-dimensional self-gravitating
fluid, see Binney \& Tremaine 1987).
Here $\kappa$ is the epicyclic frequency, which is of the order of the angular frequency
of the fluid elements (so expresses how fast they move due to rotation), $c_s$ is the thermal sound
speed, hence encapsulates the effect of thermal pressure, ad $\Sigma$ is the disk surface density,
accounting for the self-gravity of the system ($G$ is the gravitational constant).
The Toomre stability analysis shows how, under a small initial perturbation,
short wavelengths are stabilized by thermal pressure and long wavelengths are stabilized by
rotation. The system is probe to collapse if $Q < 1$, while it is stable if $Q > 1$.
The conditions in the definition of $Q$ are all local in an axisymmetric system, which implies that the
parameter can be used in any system irrespective of its global mass and internal velocity distribution.
While this is useful because it means the definition has general validity, 
it also means that when one uses $Q$ in a global disk model, a conventional
procedure in the astrophysical literature, the Toomre parameter should be viewed as a phenomenological
parameter which is not rigorous anymore and whose interpretation requires caution (often this is aided
a posteriori by numerical simulations).
Once $Q < 1$, whether local collapse occurs or not depends on the second condition, the so-called
Gammie  criterion, whereby the
radiative cooling timescale has to be comparable or shorter than the local orbital time. Local collapse, often referred to
as fragmentation, can indeed occur only if the heat gained while the gas tries to compress in order
to collapse gravitationally is removed rapidly enough by radiative cooling. The local orbital time
measures the speed at which gas is gravitationally pulled together in the attempt to collapse (formally it is
the characteristic dynamical timescale of the disk annulus with radius $R$ which $Q < 1$). If not collapse
is halted and the disk becomes stable as a result of heating ($Q >1$). Note that the heat here comes from
the action of self-gravity itself which triggers the compression, not by an external agent (this is not t
necessarily true in a disk that is externally perturbed).

There are important nonlinear aspects of self-gravitating disks that the Toomre analysis cannot capture,
some related also to the global flow properties that cannot be explained with a local approach. For example,
when $Q$ approaches $1$ from above in a disk annulus, 
shocks occur at the interface
between  the overdense regions that are becoming unstable , such as spiral arms, and the background flow.
In such a system fluid elements do not move on circular orbits as assumed for the unperturbed fluid background in the
Toomre analysis. 
As a result, the gas velocity field tends to become
chaotic from orderly even before $Q$ drops below unity, 
with its random kinetic energy growing at the expense of its initial gravitational potential and 
rotational  energy. This state is referred to as {\it gravitoturbulent}. Efficient angular momentum transport is
possible in this case (eg Binney \& Tremaine 2003) , as in any turbulent fluid. However, here, at variance with 
conventional hydrodynamical turbulence, the energy source comes from self-gravity and the injection scale of turbulence is 
associated to the wavelengths of the perturbations that can become unstable based on the Toomre analysis. Largest wavelength 
modes, such as 
global spiral patterns, have wavelengths as large as a fraction of the disk radius. This means the injection scale 
can be quite large, which is why angular momentum transport is self-gravitating disks has a global character.
As long as dissipation is sufficiently local, gravitoturbulence itself can be described 
as a source of effective viscous heating which leads to self-regulation of disks towards marginal stability (
Lin \& Pringle 1987; Lodato \& Rice 2004;2005).
Other departures from the Toomre analysis are due to the fact that disks have a finite thickness.
Numerical simulations that can follow non-linear stages of the instability and can do that in a global, 
three-dimensional set-up, find that $Q < 0.7$ rather than $Q < 1$ is necessary for fragmentation, namely a more
restrictive condition that
for the razor-thin disks assumed in the Toomre analysis (Nelson et al. 1998; Mayer et al 2004; Durisen et al. 2007). At least in 
keplerian disks simulations also find the system 
is driven quickly to fragmentation a global sustained spiral pattern develops where $Q < 1.4$ somewhere 
(Mayer et al. 2004;  Durisen et al. 2007; Kratter \& Lodato 2016).

Recently substantial work has been devoted to pin down  the critical cooling timescale for fragmentation. In recent
past different works in the context of protoplanetary disks have shown non-convergence of results with increasing 
numerical resolution, which has cast doubts on the physical understanding of the fragmentation process. The tendency
has been that the critical cooling time for fragmentation was increasing by an order of magnitude as resolution
was increasing by two orders of magnitude (e.g., Meru \& Bate 2011; 2012).
However, recently evidence has been gathered that non-convergence is likely caused by numerical inaccuracies
in capturing the strongly non-linear regime, with potential difference sources of errors coming from different
aspects of numerical schemes. One prime example is numerical viscosity, which is an explicit term in the fluid
equations  in particle-based codes such as as Smoothed Particle Hydrodynamics (SPH) but also in many popular
finite volume codes designed to model disks (Meru \& Bate 2012). 
Very recent work employing more modern numerical
techniques, such as the lagrangian mesh-less finite mass method (MFM) in the GIZMO code (Hopkins et al. 2015), 
that can follow more accurately the evolution of flow vorticity, and thus angular momentum transport,
have finally recovered convergence and determined 
that the critical cooling time for fragmentation is short, being essentially
equal to the local orbital time (Deng, Mayer \& Meru 2017). This agrees with the original results
of Gammie (2001) obtained with 2D shearing sheets. It implies fragmentation is harder to attain than previously
thought. 

In summary, based on the discussion above three-dimensional isolated disks, including protogalactic disks, could sustain efficient
angular momentum transport without fragmenting down to really low values of the Toomre parameter, $Q < 0.7$. Since protoagalactic
disks are not keplerian, whether or not $Q < 1.4$ marks  the likely fragmentation threshold in this case is unclear. Also, as time
progresses and protoagalactic disks become more metal-rich the radiative cooling timescale could shorten if gas
is optically thin, which should favour
fragmentation. In present-day galaxies, indeed the dominant neutral gas phase has a low-density and is optically thin,
and indeed Giant Molecular Clouds (GMCs) are believed to form from gas disk fragmentation (e.g., Agertz et al. 2009).
On the other end, if at high redshift disks are more gas-rich and denser, and thus become optically thick, this
would lengthen the cooling timescale, which would have the opposite effect (in this case an increasing metallicity
would favour stability even further as it would increase gas opacities,which in an optically thick system results
in longer cooling time).  The properties of the interstellar medium are thus crucial in this context.
We will revisit this aspect in the next two sections, in which we will also drop the assumption of an isolated,
unperturbed disk.

\begin{figure*}[]
\includegraphics[width=0.9\textwidth]{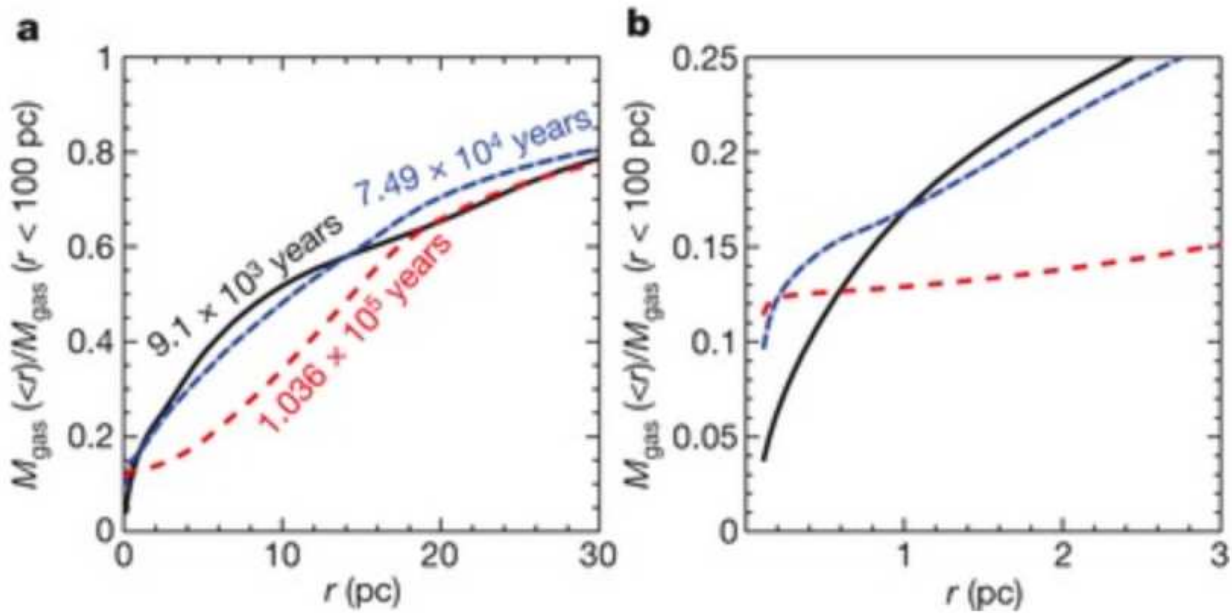}
\caption{The figure shows the growth of the central gas mass distribution following the
multi-scale galaxy merger in the MA10 paper, at tens of parsecs (left) and at parsec scales
(right). Different curves correspond to different times starting from the completion of the
galaxy merger as indicated in the labels. Adapted from MA10. }
\end{figure*}

\section{Simulations of the merger-driven direct collapse scenario}

The merger driven model has been investigated with simulations of binary mergers between massive gas-rich disks
whose structure is similar to present-day spirals (Kazantzidis et al. 2005; Mayer et al. 2010,2015, hereafter MA10 and MA15,
respectively). 
For the simulations we employed the SPH codes GASOLINE and GASOLINE2 (Wadsley et al. 2004; Keller et al. 2014).
The choice of modeling galaxies based on prexent-day disks is a simplification 
as we have observational evidence
that, already at $z \sim 2-3$, massive gas-rich galaxies appear much more irregular in structure and turbulent relative
A turbulent disk could boost certain processes occurring in gas flows
that are crucial to achieve the prominent multi-scale inflows of our scenario. We will discuss these aspects later on
in section 7 along with the prospects of incorporating them by starting directly from cosmological simulations of 
massive galaxies.
New realistic cosmological hydro simulations of $z \sim 10$ 
mergers in highly biased regions are indeed in progress (Capelo, Mayer et al., in preparation). Among
the simulations published so far the earliest ones adopted simplified thermodynamics, employing
a fixed effective equation of state, while more recent ones included radiative cooling and heating explicitly.
In the next two subsections we describe both types of simulations. 

All our published simulations adopted the same initial conditions for the model galaxies. 
The two galaxy models employed in the simulations of major galaxy mergers 
represent identical multi-component systems with virial masses of $M_{vir} 
= 10^{12} M{\odot}$. They are 
constructed following the method by Hernquist (1993) and choosing 
the structural parameters in agreement with the scaling laws predicted by the $\Lambda$CDM model (Mo et al. 1998; Kazantzidis et al. 
2005). The baryonic disk is embedded in a  dark matter halo that follows 
the Navarro-Frenk-White (Navarro et al. 1996) profile with concentration $c = 12$. The stellar and gaseous disk follow the same
exponential density distribution, and a stellar bulge is also included.
The stellar disk has a mass of $M_{disk} = 6 \times 10^{10} M_{\odot}$, a scale radius of $R_{disk} = 3.5$ kpc, and a 
scale height of 
$z_{disk} = 350$ pc. The gaseous disk has the same scale lengths as the stellar disk. The gas mass, which is gradually 
reduced
by star formation,  is still about 
$10 \%$ of the disk mass at 
the time of the 
last pericenter passage. The bulge is a Hernquist (1990) model with $1/5$ of the disk mass 
and a scale length of  0.7 kpc. We 
refer to MA10 
for more details on the numerical parameters of the large-scale merger simulation (see also Kazantzidis et al. 2005; Mayer et al. 2007).
We note that the adopted virial mass is comparable to that of a Milky-Way-sized galaxy at z = 0, and it 
corresponds to $\sim 5\sigma$ dark matter fluctuations at $z = 8-9$ in the WMAP9 cosmology 
(the cosmology assumed in the simulations).
With the assigned disk and bulge mass the galaxy is consistent with the stellar-to-halo mass relation constrained by abundance
matching and other statistical measurements correlating galaxy and halo properties up to very high
redshift (Behroozi et al. 2013; Moster et al. 2017),
although we caution that no robust measurements exist at $z > 7$.
The galaxies are placed on a parabolic orbit whose parameters are chosen to be typical of halo/galaxy pairs in large cosmological
volumes (Kochfar \& Burkert 2006).

In order to follow the gas inflow to sub-pc scales we had to increase the resolution of the hydrodynamical merger
simulations in subsequent steps.
We split individual gas particles within a spherical volume of radius 30 kpc into eight child particles in two
subsequent steps. We correspondingly reduce their gravitational softening, 
achieving a final resolution of $\sim 2600 M_{\odot}$ and $0.1$ pc. 
Pre-existing star particles and dark matter particles are not split, and their softening is left unchanged. They essentially provide a smooth background potential to avoid spurious two-body heating against the much lighter gas particles. The momentum conserving splitting procedure, 
is described in Roskar et al. (2015). Note that in the earlier, non-radiative simulations of MA10 a simpler
splitting algorithm exploiting only a {\it gather} SPH kernel operation was employed . Later we have found out
that using a {\it scatter} operation is numerically more robust when re-assigning internal energies to child particles in presence
of steep temperature gradients (in MA10 this was not important as the use of an effective equation of state was
leading by construction to weak local temperature gradients). We point the reader to Roskar et al (2015) for
a detailed description of the updated particle splitting algorithm.
The choices of when to split the gas particles and the size of the 
splitting volume are the same in both MA10 and MA15 and are discussed 
in the Supplementary Information of MA10. They are motivated by inducing minimal numerical fluctuations. To this end the refined region 
was large enough to avoid any contamination of low-res particles for the entire duration of the simulations. 

After the two galaxy cores reached a separation $<100$ pc, most of the gaseous mass is collected in the inner 100 pc volume and it is traced 
with  more than a million gas particles. By performing numerical tests, we have verified that, owing to the fact that gas dominates 
the mass and dynamics of the nuclear region, the large softening adopted for the dark matter particles does 
not significantly affect the density profile of the inner dark halo that surrounds the nuclear disk. 
These and other numerical tests showing the robustness of the technique were presented in Mayer et al. (2007).

\begin{figure*}[]
\includegraphics[width=0.9\textwidth]{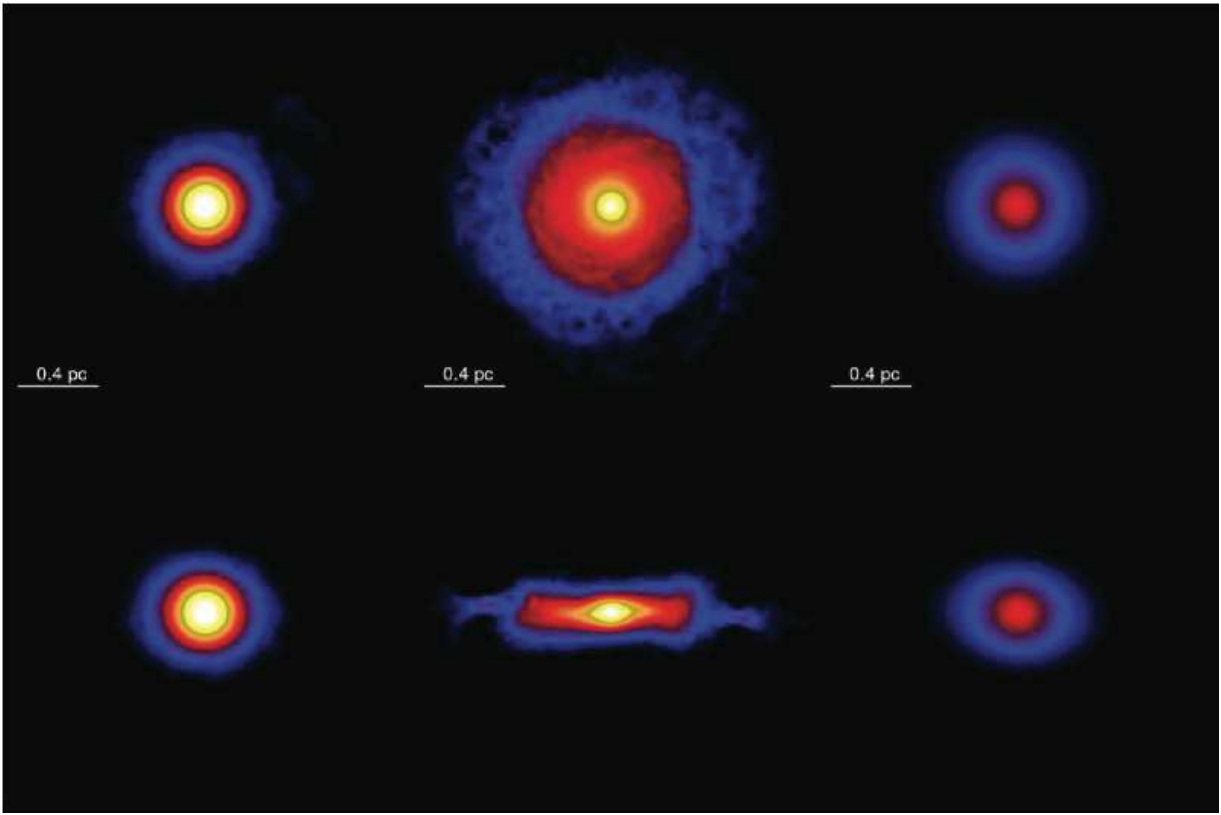}
\caption{The structure of the final supermassive cloud at pc scales in the
MA10 simulations adopting an effective EOS. We show the color coded density
maps for the face-on (top, along the angular momentum vector) and edge-on (bottom, perpendicular
to the angular momentum vector) views of the cloud after
$10^5$ yr from the end of the galaxy merger. From left to right, the end states for equal-mass mergers with
galaxies of of different mass are shown (see section 4.1). Note the lowest mass case does not produce a gravitationally
bound cloud, while the other two do. The reference case, for a virial mass of $10^{12} M_{\odot}$, is
shown on the left. The intermediate mass case gives rise to a rapidly rotating cloud which an unstable
to global non-axisymmetric modes that could continue the inflow to smaller scales (see also section 5).}
\end{figure*}

\subsection{Non-radiative simulations}

In MA10 we simulated multi-scale gas inflows in equal mass mergers between two identical galaxies modelled
with the initial conditions described in the previous section. The SPH code GASOLINE was used to model the combined hydrodynamical
and gravitational physics, while radiation was not treated explicitly, as we explain below. Star formation and thermal feedback
from Supernovae Type II were included only in the large scale merger simulation but not in the re-sampled simulation stage
after particle splitting was applied (see MA10 for details).
We also performed two additional simulations of equal mass merging galaxies
in which the mass of the two galaxies was reduced by a factor of 5 and 20, respectively. These addition models were constructed by
simply scaling the radius and velocity in order to preserve the same density and virial equilibrium condition. This is consistent
with the notion that, for a given cosmology and redshift, the critical overdensity for collapse of dissipationsless cold dark
matter is fixed (Mo et al. 1998).
This yields  scaling relations between virial mass $M_{vir}$, virial radius $R_{vir}$ and virial velocity
$V_{vir}$ in the form $M_{vir} \sim R_{vir}^{1/3}$ and $V_{vir} \sim R_{vir}$. Central in the MA10 work was the two-step procedure in running
the simulation to capture sub-pc scale dynamics while starting from a galaxy merger in which the initial dark matter halos extended
out to hundreds of kpc.  The other central aspect was the adoption of simplified thermodynamics of the gas. We employed
an adiabatic equation of state with variable adiabatic index $\gamma$, of the form $P = (\gamma - 1) \rho u$, where $P$ is the
thermal pressure, $u$ is the internal energy (the gas is assumed to be ideal), and $\rho$ is the density. The range of $\gamma$ values
adopted was determined based on the density range in the simulations, and was  based on a simple equilibrium model of the ISM 
illuminated
by a ionizing flux produced by local star formation (Spaans \& Silk 2000).
The internal energy equation was solved, thus capturing irreversible heating from shocks,
which is important in galaxy mergers.

In the first stage of the mergers dynamical friction between the two extended dark matter halos erodes the orbital energy of the two galaxies, which 
sink towards one another and merge in less than a Gyr. When the two galaxies are at a distance of about 5 kpc we applied the particle
splitting technique and decrease the gravitational softening in two subsequent steps (from 100 pc to 5, and then to  0.1 pc).
The final collision of the two galactic cores produces a massive turbulent rotating nuclear disk with a mass 
of about two billion solar masses  and a radius of about 80 pc. 
The disk is born in an unstable configuration, with a prominent two-armed spiral pattern imprinted by the collision and sustained by 
its own strong self-gravity. The gas has a high turbulent velocity ($100$ km/s),whose gravitourbulent origin we discuss
in the next section, and rotates at a speed of several hundred kilometres per second within 50 pc. The disk 
is stable against fragmentation but the strong spiral pattern swiftly transports mass inward and angular momentum outward 
About 10,000 years after the merger is completed more than 20\% of the disk mass has already accumulated within 
the central few parsecs (Figure 1) where the inflow rate peaks at over $10^4 M_{\odot}$  per year, 
orders of magnitude above inflow rates in isolated protogalaxies (see sections 1-2).

The gas funnelled to the central region (2-3 parsecs) of the nuclear disk settles into a rotating, pressure-supported cloud. The 
density of the cloud continuously increases as it gains mass from the inflow until it becomes Jeans unstable 
and collapses to sub-parsec scales on the local dynamical time of 1000 years (see Figure 2). 
The supermassive cloud contains around $13\%$ of the disk mass (Figure 1). 
The simulation is stopped once the central cloud has contracted to a size comparable to the spatial resolution limit. At this point the cloud is still Jeans unstable. With greater resolution its collapse should continue because the equation of state would become essentially isothermal at even higher densities. With its steep density profile, a power law of density with exponent steeper than $-2$,  
at radii as small as a thousandth of a parsec the cloud would be as massive and dense as a protostellar envelope that can 
collapse directly into a massive black hole seed embedded in a gaseous envelope, the so-called quasi-star proposed by Begelman 
and collaborators (Begelman et al. 2006; Begelman \& Volonteri 2010). Other routes, such a supermassive star which later collapses
into a massive black hole encompassing a large fraction of its mass, or even direct formation of a supermassive black hole via
the relativistic radial instability, are also possible. We will discuss pathways more in detail 
in section 5 using the newer set of radiative simulations.

\begin{figure}[]
\includegraphics[width=0.9\columnwidth]{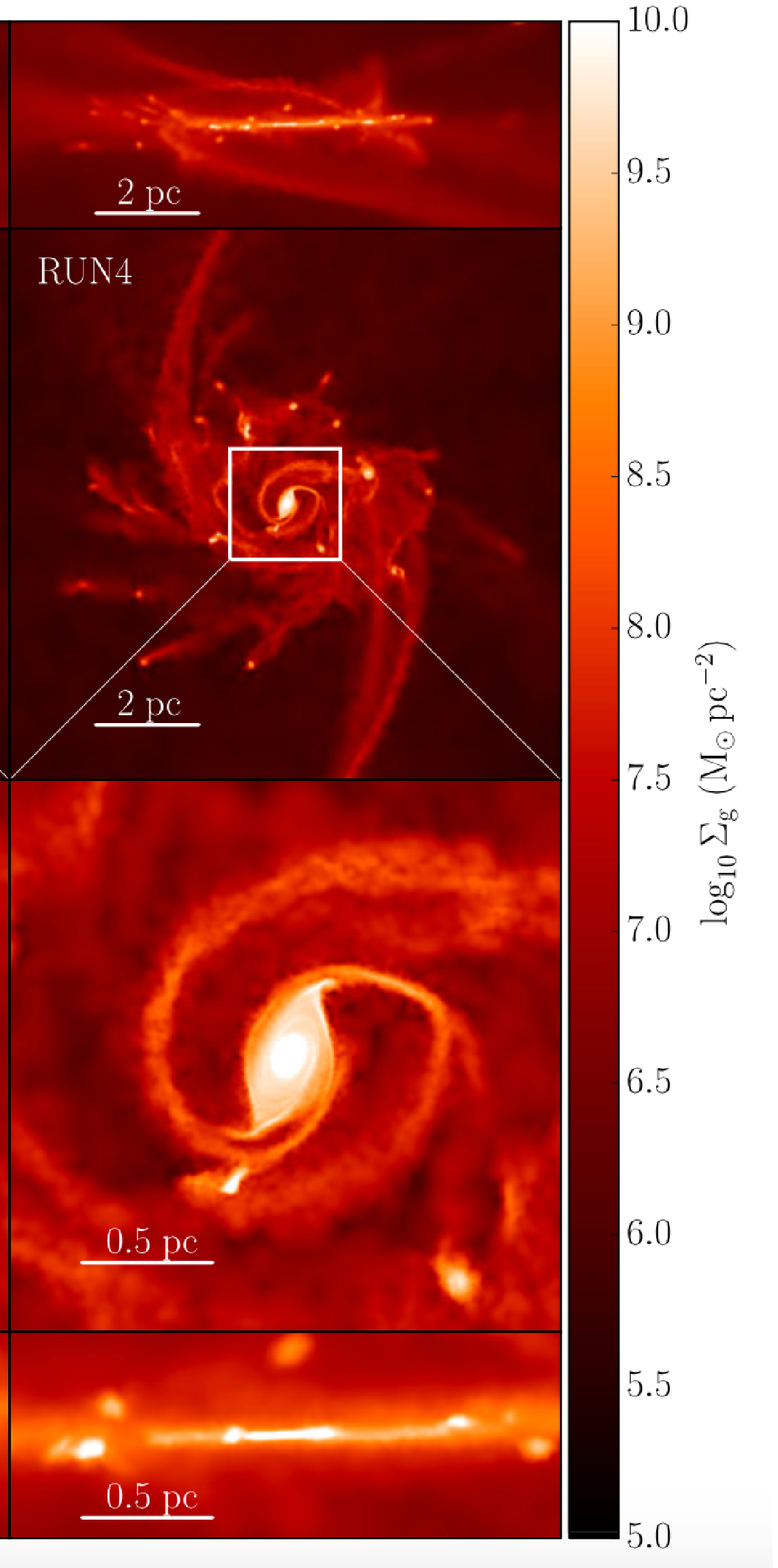}
\caption{Face-on and edge-on projected gas density maps of the nuclear region of the merger at $t_0 + 5$ kyr ($t_0$ is the time 
corresponding to the merger of the two central cores.)
showing a disk-like object with 
radius 5 pc (first and second panel are edge-on and face-on, respectively), and a compact inner disk-like core less than a parsec in size (third 
and fourth panels are face-on and edge-on, respectively). We show one particular run from the MA15 paper, RUN4, which employed a modern
version of the SPH hydro solver using pressure-energy formulation (GDSPH), thermal and metal diffusion, and a Wedland kernel. As seen
in MA15, the structure of the turbulent self-gravitating mini-disk is almost independent on SPH solver and details of the
cooling model, which all vary in the different runs presented in MA15.}
\end{figure}

The formation of the collapsing cloud occurs in only $10^5$ years after the completion of the merger, 
a timescale much shorter than the $10^8$ years needed to convert most of the nuclear gas into stars during the starburst that
accompanies the merger. From such earlier simulations it was already clear that with such short formation timescales for the
massive BH precursor object one overcomes a major problem of the more conventional direct collapse
scenarios, namely the need to suppress cooling and star formation. Simply gas inflows occur on a timescale
shorter than the typical gas consumption timescale into stars. This point will be strengthened further 
with the radiative simulations described in the next section, in which we will see that, additionally,
the very inner  dense pc-scale region is completely stable to fragmentation and star formation even when all 
cooling channels are active. The high inflow rates measured suggest the simulated system belongs to the
{\it heavily mass loaded} case studied by the analytical model presented in section 3, where stability
was predicted irrespective of cooling. This will imply that an effective EOS approach should suffice, However
it needs verification with fully radiative simulations, which will be the subject of the  next section.

The additional simulations with galaxies with masses a few times $10^{11} M_{\odot}$ show that in this case 
even stronger gas inflows can arise (Figure 2).
This can be understood in terms of the factors that determine the spiral instability in the nuclear disk, namely 
its density and temperature. At a few times   $10^{11}$ the gas mass in the disk is still of order $10^{10} M_{\odot}$
prominent enough to produce a dense self-gravitating nuclear disk after the galaxy merger, but such a disk is colder 
and thus even more unstable to spiral modes than is our reference simulation. 
Therefore our direct 
collapse mechanism could have produced massive black-hole seeds in galaxies that were more common than the high-$\sigma$ peaks hosting the 
bright quasars at $z> 6$, including the most massive progenitors of our own Milky Way. 
On the contrary, we find that in merging galaxies with virial masses of only $5 \times 10^{10} M_{\odot}$, similar
to the present-day Large Magellanic Cloud, a satellite galaxy of our Milky Way,  the inflow is too weak to 
trigger the gravitational collapse 
of the central region (Figure 2).
The light nuclear disk that forms has a very low gas surface density that cannot support a strong 
spiral instability.

\begin{figure}[]
\includegraphics[width=1\columnwidth]{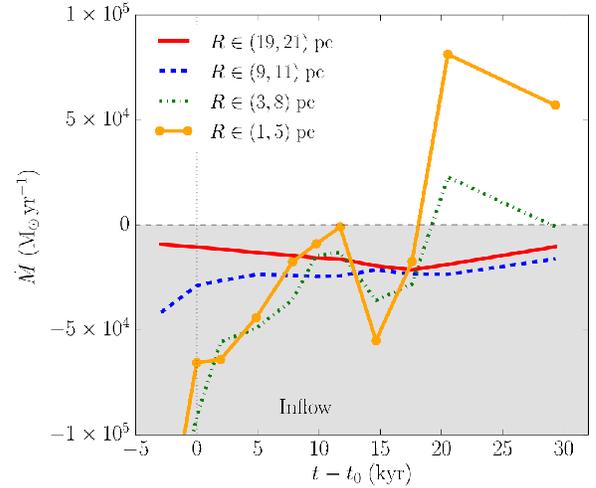}
\caption{Time evolution of the gas accretion rate at different radii from the center of the merger remnant. 
The accretion is computed inside 
cylindrical shells of inner and outer radii marked in the legend and vertical thickness 2 pc. RUN4 from MA15 is used, which is also
shown in Figure 3. Adapted from MA15}
\end{figure}

The trend of strength of the gas inflow versus galaxy mass can be understood as a consequence of 
how the two critical factors determining the stability of the nuclear disk, effective
pressure/temperature (including the turbulent part) and surface density, scale with galaxy
mass. The argument goes as follows (in the remainder we will always use the scaling laws
introduced by Mo, Mao \& White 1998). Thermal and non-thermal (turbulent) energy content
of the disk arising from the merger will be determined by the kinetic energy injected
by the shock and then thermalized after the two galactic cores collide (note that because
of our choice of a stiff EOS ($\gamma < 7/5$) radiative losses are implicitly small, hence it is
conceivable to assume thermalization of the kinetic energy). The energy imparted by the
shock is $E_{shock} \propto M_{vir} V_{vir}^2$ (where the product of the galaxies' virial mass and virial
circular velocity squared measures the orbital kinetic energy lost in the collision, assuming
a circular orbit for simplicity) , or, since $V_{vir} \sim M^{1/3}$ , we have $E_{shock} \sim M_{vir}^ {5/3}$. On the
other hand, the surface density of the nuclear disk, which is proportional to that of the host
galaxy merger remnant, scales as $\Sigma \sim M_{vir}/R{vir}^2 \sim M_{vir}^{1/3}$. The disk angular 
frequency
$\Omega \sim V_{vir}/R_{vir}$ does not depend on $M_{vir}$ since $V_{vir} \sim R_{vir}$. Therefore, if we define the effective
sound speed in the nuclear disk as $c_s* \sim E_{i}^{1/2}$ (where $E_i$ is the internal energy of the gas,
including both the thermal and the kinetic energy component), and assume 
$E_i \sim E_{shock}$, 
namely that shock heating sets the level of internal energy of the nuclear disk arising after the collision, the Toomre parameter of 
the nuclear disk is $Q \sim c_s / M_{vir}^{1/2}$. In other words, nuclear disks in higher mass galaxies 
are increasingly more stable, as the energy imparted by the collision increases more rapidly than the disk self-gravity.

\begin{figure*}[]
\includegraphics[width=0.9\textwidth]{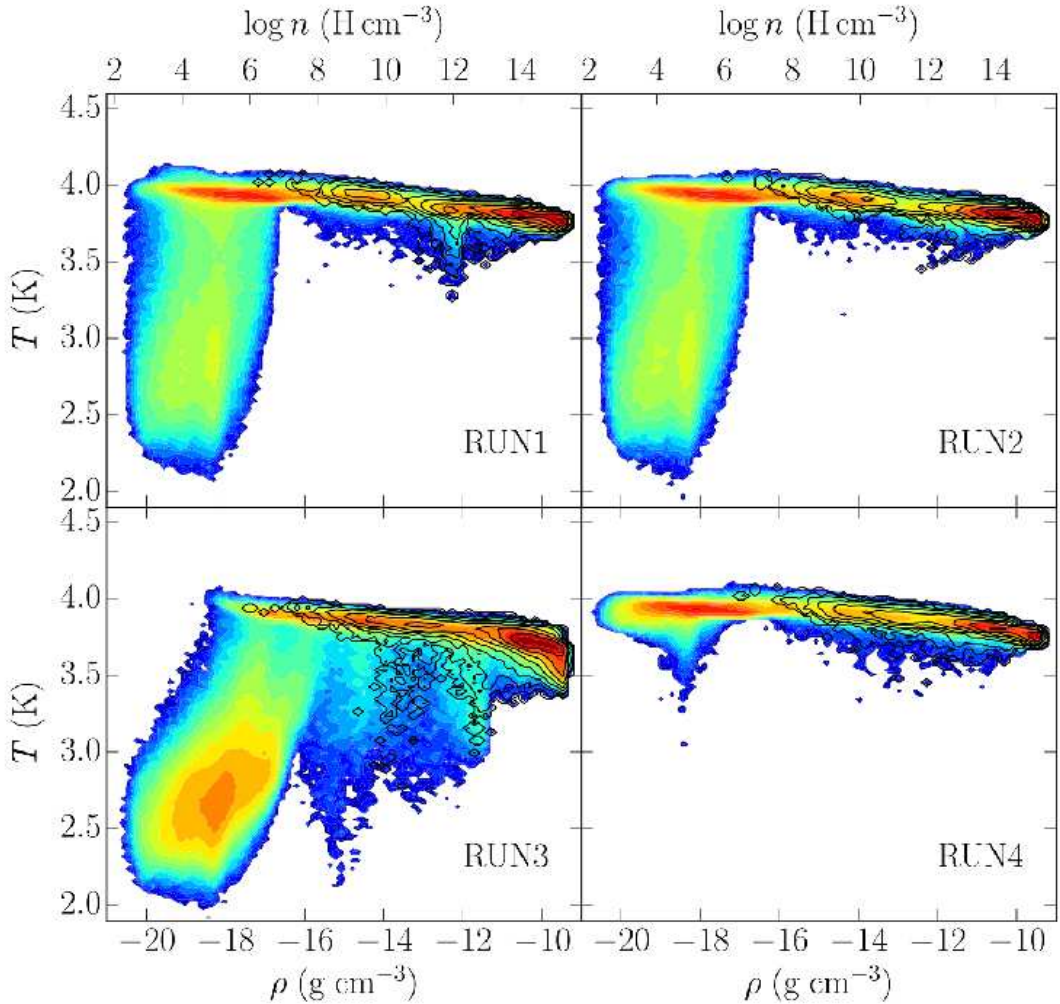}
\caption{Phase diagrams of the particles within a 50 pc volume at $t = t_0 + 5$ kyr for the radiative simulations
of MA15. From top-left to 
bottom-right, we show results for different runs presented in MA15. The two shown at the top are from runs that used vanilla SPH and metal-line 
cooling but
do not include the improved treatment of cooling, heating and self-shielding at high optical depths in the dense core. They differ
for details in the implementation of star formation (see MA15). The two shown at the bottom use vanilla SPH but include the effective
temperature-density model for high optical depth regions in one case (left) , or the modern GDSPH implementation with diffusion terms
and Wedland kernel, yet with only standard metal-line cooling using CLOUDY (RUN4, on the right). Black contours are superimposed to show the 
particles that are located in the  inner parsec region, within the inner compact disk. Note that this latter region has almost the same phase
diagram in all four runs, consistent with the expectations from the analytical model described in section 3 for which the details of the
cooling should not matter rather the behaviour is determined by the gas inflow/accretion rate from large scales.
Adapted from MA15.}
\end{figure*}

This is the first key result, which explains why the intermediate mass merger produces a more unstable nuclear disk, 
and thus a more massive collapsing central gas cloud compared to the reference simulation with higher mass galaxies. 
However, for increasingly lower mass galaxies the energy imparted by the shock, 
which scales as $M_{vir}^{5/3}$, will eventually become 
lower than the initial internal energy of the colliding galaxy cores, so that the surface density of the 
nuclear disk becomes the most important parameter. 
In this case we can assume that $c_s$ does not depend on $M_{vir}$ anymore , so that $Q \sim \Sigma^{-1} \sim M_{vir}^{-1/3}$ 
hence lighter disks are less dense, and therefore more stable. 
This explains the increased stability of the lowest mass merger 
that we considered, which results in its inability to undergo central Jeans collapse.
The role of shock heating in setting the boundary condition for temperature in the nuclear disk will 
be confirmed in the radiative simulations of the next  section.

As a final cautionary remark,note that the inclusion of radiative cooling, star formation and feedback processes may invalidate
the simple scaling arguments employed here. These hold because in EOS simulations the nuclear disk is relatively homogeneous in its temperature 
structure so that a near steady-state can be assumed. Yet the stifling of the multi-scale instability at low masses should remain as it is generally 
agreed radiative and kinetic feedback from stars and supernovae becomes a dominant effect eventually, and generates primarily outflows
rather than inflows. Indeed, a large number of dwarf galaxy formation simulations (Governato et al. 2010;
Onorbe et al. 2015; di Cintio et al. 2014; Wetzel et al. 2015) show that, in  galaxies with halo masses only a few 
tens of billions of  
solar masses, supernovae-driven gas outflows prevail over merger-driven gas inflows, so that
no central gas concentration results and no nuclear disk forms. This is indeed the widely
accepted explanation of why dwarf galaxies are bulgeless while massive galaxies always grow a
central bulge (Governato et al. 2010).
These inferences are based
on studying galaxy formation models at relatively late stages ($z < 3-4$), which could cast
some doubts on their general validity since at higher redshift gas accretion from the cosmic
web should be more prominent and characteristic densities of objects are much higher. However,
some recent works studying low mass galaxies ($M_{vir} < 10^{11} M_{\odot}$) at $z > 5$ in high-resolution hydrodynamical
simulations suggest that feedback is also more intense when the specific star formation rate
(the star formation rate per unit surface area) is high enough to match that inferred 
from the data. Feedback then stifles the formation of any coherent dense structure in the inner few tens
of parsecs (Fiacconi et al. 2017). While more investigations of the early stages of the formation
of low mass galaxies are required, current evidence from galaxy formation simulations strengthens
our conclusion that formation of massive black-hole seeds by
our flavour of direct collapse would only occur in massive galaxies.

\subsection{Radiative simulations}

\subsection{The radiative cooling model}

Starting from the same initial conditions described above for the more massive models (with halo virial mass
of $10^{12} M_{\odot}$)
we have run four different simulations
using GASOLINE2, an updated version of the GASOLINE code (Keller et al. 2014). Individual runs differ in the sub-grid 
model for radiative cooling as well as for the specific implementation of the SPH equations of hydrodynamics.
In all of the runs, we adopt the metal-dependent, optically thin cooling introduced in Shen et al. (2010). It considers 
tabulated cooling rates in ionization equilibrium, while for H and He we directly compute the rates without assuming equilibrium 
(Wadsley et al. 2004). As in MA10, we assume solar metallicity gas, consistent with observational constraints on the metallicity of 
the hosts of high-z QSOs (Walter et al. 2004). In one run we include a treatment of the optically thick 
regime, thereby gas above a density of $n_g = 0.1$ cm$^{-3}$ is on an  
equilibrium temperature-density relation calibrated on 2D 
radiative  transfer calculations for a nuclear starburst model. The latter is 
based on an improved version of the Spaans \& Silk (2000) model (see Roskar et al. 
(2015) for details). This "thermal balance model accounts for self-shielding effects in a dense ISM
and for the heating processes that operate at very high density: (1) 
cooling by molecular lines and by collisions between dust and molecules with metallicity-dependent opacity effects due to absorption 
and scattering of photons by dust; (2) IR dust radiation; (3) photoelectric effect on dust; (4) atomic 
and molecular line trapping 
in an ISM irradiated by stellar light; and finally (5) heating by cosmic rays. 
Cooling by $H_2$ is not taken into account in any regime because it is superseded by other molecular 
cooling channels in a metal-enriched gas, 
such as CO, which are  included in our "metal cooling" CLOUDY module. Furthermore, 
cooling by fine structure metal lines is actually the dominant process, providing almost two orders of 
magnitude higher energy loss rate per unit volume compared to molecular cooling at a temperature of a few thousand K
even in the high density nuclear region of the merger remnant (see MA15).

The "thermal balance model" assumes the presence of a uniform UV photon radiation field produced by a 
starburst. Indeed in the large scale merger simulations, when the galaxies are still in the process of coalescing, a 
starburst with a strength of $\sim 100 
M_{\odot}$ / yr takes place in the inner kiloparsec (see MA10). The latter star formation rate thus 
assumed to  determine the stellar UV flux boundary condition in the thermal balance model. 
While the intensity of the starburst is actually a free parameter, the assumed value is on the low side. 
Indeed high resdhift starbursts can have easily an order of magnitude higher intensity. This would maintain a higher 
equilibrium temperature , resulting in even stronger stability of the nuclear compact disk against fragmentation.
Note that in this sense a similar notion as in the conventional direct collapse scenario in protogalaxies,
namely an external UV flux, would seem to be apply to our model too. However, we will see that, while 
heating by a starburst is inevitable in our framework, and not an ad-hoc assumption, results will not
depend on that or on including self-shielding effects to reduce the cooling rate.

In order to avoid artificial fragmentation we enforce a 
a pressure floor so that  the Jeans mass is resolved locally by at least one resolution element, namely
one SPH kernel (see  MA15
and Roskar et al. 2015).
Most importantly, we do include star formation and blast-wave supernova feedback. The implementation 
of feedback in the code follows closely Stinson et al. (2006); stars 
form from gas following a Schmidt law in regions above the density threshold 
$n_{th} = 10^4$ H cm$^{-3}$ and below the temperature  threshold $T_{th}$ = 500 K,
provided that they are in a convergent flow (measured by the local divergence of the
gas velocity field).

We also compare runs with different implementations of SPH that improve on the issue of lack of mixing
due to the tensile numerical instability  (Agertz et al. 2007), allowing
to  modelling  multi-phase turbulent flows. In particular we compare vanilla-SPH, using a standard 
density-energy formulation, and GDPSH, which uses a formulation of the hydro force similar to the 
pressure-energy approach in Hopkins (2013), and includes also a thermal and metal diffusion term (see MA15 
and references therein as well as Shen et al. 2010). The results of the selected run shown 
in Figures 3,4,5 and 6 are those adopting GDSPH.

\subsection{Enhanced inflows and warm loaded nuclear disks}

The  main result of these simulations is that, when radiative cooling is included, the multi-scale gas inflows are even more
prominent than in the earlier,more idealized simulations with an effective equation of state performed in MA10. This is opposite 
to the naive expectation that radiative cooling, which here includes metal lines as the gas is assumed by construction to be
of solar metallicity, would lead to widespread fragmentation and the disruption of the central inflow. Fragmentation does happen
but it is confined only to the outer part of the nuclear compact disk that forms rapidly (Figure 3), in less than $10^4$ yr, at 
the center
of the merger remnant. The disk encompasses more than $10^9 M_{\odot}$ within 1 pc only a few kyr after it begins to
assemble (Figure 6).
Figure 3 shows the state of the system after at two different scales,
the smallest being about a parsec (equivalent to 10 resolution lengths). The stronger inflow can be measured by the gas infall rates
at various scales (Figure 4), which peak at a value almost an order of magnitude higher than in MA10.This is because efficient cooling
renders the inflow nearly isothermal in the nuclear region, removing efficiently the heating generated by shocks.
The temperature distribution can be appreciated by inspecting the phase diagrams (Figure 5).
As a result a much compact,
denser and thinner disk forms relative to the MA10 simulations; its size is below 10 pc as opposed to 100 pc in the previous study.
Fragmentation is absent in the inner dense disk core, because the temperature hovers around $5000-6000$ K
rather than dropping significantly below that (compare Figure 3 and Figure 5).
Therefore,somehow unexpectedly the thermal state of the gas at pc scales is very similar to
what is obtained in simulations of metal-free protogalaxies illuminated by a dissociating UV source, as in the more conventional
direct collapse scenario. 

The fact that the inflow leads ultimately to a relatively warm nearly isothermal central collapse in a metal-enriched gas is not
a trivial outcome. It
can be considered the second important result of our work. It is a robust outcome as it is obtained irrespectively
of the cooling model adopted in the simulations, for example with or without the thermal balance model, and
also does not depend on the specific SPH variant employed (Figure 5).

In order to understand the origin of the nuclear temperature we need to consider the timescales
of the key processes involved, and the fact that the system transitions from optically thin to optically thick at different scales
as the density increases towards the center. The key underlying point is that there is a competition between radiative cooling and
heating by shocks and turbulence. Shock heating dominates in the infalling gas where Mach numbers can be higher than 10 (see MA15),
while once gas settles in the dense disk turbulence resulting from gravitational unstabilities in the disk itself becomes the
dominant heating source. We verified that, in
the absence of any radiative cooling, with the large supersonic 
infall velocity found in the simulations ($\sim 1000$ km/s) the flow would undergo a strong shock
yielding a post-shock temperature $> 10^6$ K. The infalling gas, though, is still at an optical depth
low enough for cooling to take over, bringing the gas to a temperature of few thousand kelvin, but
not below that (the shock becomes essentially isothermal).
Indeed  we verified that the combination of high densities and very 
high infall velocities in the inner parsec region produces a compressional heating rate that is roughly equal 
to the cooling rate by fine structure lines around a few 1000 K, while line cooling dominates heating at temperatures above $10^4 
K$. These thermal balance arguments explain why the temperature of the disky core hovers around a few thousand K.

Within the dense disk gas cannot cool below a few thousand K because the optical depth is very large. 
We will return to the latter timescale in the next section, using the specific
conditions found in the simulations.
The phase diagrams also show that the "metal cooling" is effective at lowering the temperature in the low density, 
low Mach number regions far from 
the core (Figure 5).
We conclude that the warm temperature maintained by the central disk-like core is a very
robust, inevitable product of the supersonic gravitational infall 
enabled by the larger scale dynamics of a major galaxy merger. The merger essentially provides the
ideal case for the {\it loaded disk} model presented in the next section.

We also note that no star formation occurs in the warm core simply because the gas 
never  cools enough to meet the temperature condition of our star formation prescription. 
Sporadic star formation can occur in the densest gas pockets 
further away from the center, where metal-line cooling is effective (Figure 3).
The fact that no star formation occurs in the core of the nuclear disk is not only a direct
consequence of our star formation prescription, in which a minimum density, a maximum
temperature, and the requirement that gas is in a convergent flow locally 
are set as necessary conditions.  Indeed, even if we had set the relevant parameters to
very different values, or even if we had used a different prescription based on other
relevant properties of the ISM, such as the local Mach number, it would remain true that 
the gas in the core has a temperature well above the dissociation temperature of molecular
hydrogen (2000 K). Since it is almost universally agreed that the presence of a molecular hydrogen
phase is necessary for stars to form, it follows that, unless the thermodynamics can be
altered significantly,  no star formation should occur in the core of the merger remnant.
Likewise, the fact that some star formation occurs at the boundary of the disk, inside the dense
clumps visible in Figure 3, is expected based on the same argument as the temperature lowers
below the dissociation threshold. The dissociation threshold 
can be considered as a physically motivated temperature threshold that is more conservative
than the threshold assumed in our sub-grid model.

The simulations also address the role of turbulent kinetic energy in disk stability, an aspect that is hard to incorporate
in an analytical toy-model such as that presented in the next section 4.
At pc scales and below the inner disk gas exhibits a velocity dispersion that is higher than its thermal sound speed. This is the
typical signature of a gravoturbulent system, which is expected to settle in a state of marginal stability against fragmentation,
namely with a Toomre stability parameter of order unity (see also Choi et al. (2013;2015) and Latif et al. (2013). Indeed this is 
what is found by measuring the Toomre parameter of the disk in the simulation as $Q= v_{turb} \kappa/ \pi G \Sigma$, namely using
the turbulent velocity scale  rather than the thermal sound speed as the characteristic velocity scale as it is the largest 
of the two. Figure 7 shows that most often $Q > 1$  except in a few locations where isolated clumps do arise from fragmentation.
One exception is the
very late stage of the evolution, at $t > 10^4$ yr, when a second massive clump formed in the outer disk grows significantly in mass
and then migrates towards the central disk opening the possibility that two rather just one massive BH seed could be formed (see
section 5 and  7). Note that, given the short timescales considered here, less than $10^4$ yr, these gas clumps cannot form
stars but, at most,could collapse below the resolution into a stage similar to a protostellar phase. In doing so they would 
become emitters of infrared and optical radiation, which will have a  negligible effect on the surrounding medium as the
effective temperature of such radiation will not be higher than the equilibrium nuclear temperature of a few thousand K.

\begin{figure*}[]
\includegraphics[width=0.88\textwidth]{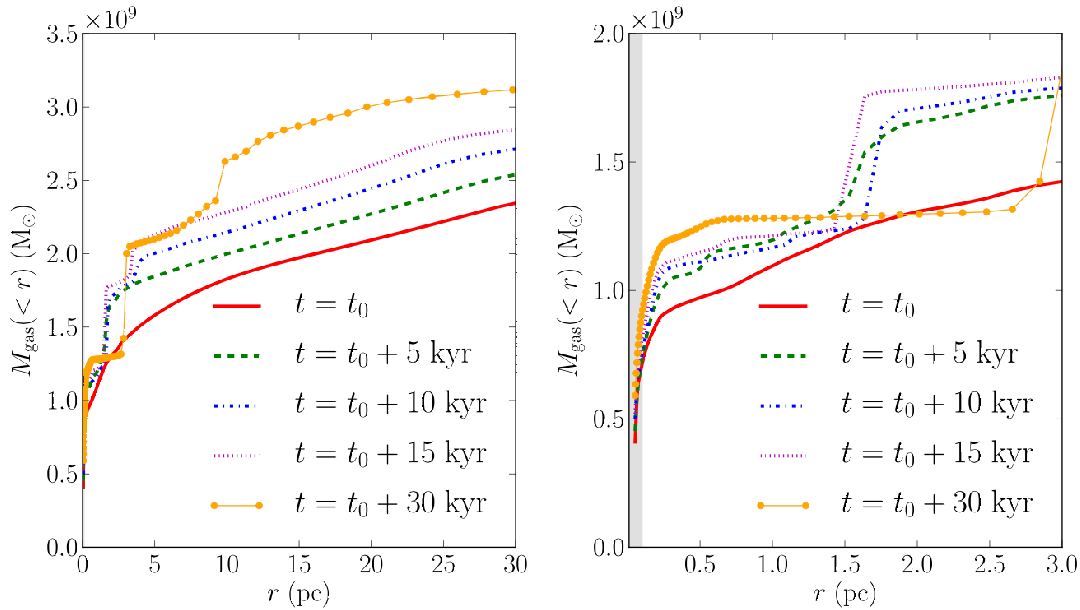}
\caption{
 Time evolution of the enclosed gas mass inside 30 pc (left) and inside 3 pc (right) for RUN4. The reference time t0 is the time of the 
coalescence of the two merging cores (see the text for details). The gray stripe in the right panel highlights the resolution limit set by the 
gravitational softening. Adapted from MA15.}

\end{figure*}

The heating resulting by gravitational instability is caused by shocks induced by spiral arms,and the turbulent 
velocity that
we measure is nothing else than a proxy for how strong is the non-axisymmetric distortion induced by self-gravity,and hence correlates
with the shock strength (see the detailed model of Boley et al. 2010 which performs a relevant analysis for self-gravitating
protoplanetary disks where the Mach number of the flow can be directly associated with the overdensity in the spiral arms). This
is why we can interpret the relative stability of the disk as the result of heating by gravitoturbulence, which keeps $Q$ generally
above unity. Note that when $Q > 1$ the cooling timescale becomes a secondary aspect of the physics. The simplest
interpretation of our results would then be that gravitoturbulence 
is the process that appears capable to self-regulate the disk into a marginally stable state
characterized by nearly isothermal conditions. Gravitoturbulence will be parametrized by an effective $\alpha$ viscosity
in the loaded disk model described in the next section. 
Furthermore, the temperature remains close
to the boundary temperature generated earlier by the supersonic infall. This will justify the 
the assumption of steady-state of the 
{\it loaded disk mode} introduced in the next section, and is consistent with the
notion that the large scale and sub-disk scale inflow rates should be  comparable. 
A difference with the loaded disk model of section 4, though, is  that in the simulations the turbulent kinetic energy is not fully thermalized
on a disk orbital time, otherwise disk temperature will increase and turbulent velocity will decrease, which is
not observed.  

There is, however, a possible major caveat in the interpretation so far. Indeed the simulations do no include
radiative transfer, hence one may worry that the thermal balance is not correctly computed in the simulations,
which could invalidate the whole argument about the self-regulated, gravitoturbulent state. 
More in general, how can we then assert that what we found should be a generic outcome of gas-rich major mergers at high-z rather than a fluke of 
the particular assumptions on gas metallicity and the cooling model in the more uncertain regime of high optical depths?
The answer is found by revisiting the diffusion timescale argument.
Let us now consider the cooling timescale in the central disk
assuming the diffusion approximation,which is valid in the very high optical depth regime within the central pc.
 Indeed, the typical optical depth in the disky core within 1 pc is ${\tau }_{\mathrm{es}}\sim {N}_{{\rm{g}}}{\sigma 
}_{{\rm{T}}}\sim {10}^{4}$, where $\sigma_T \sim 6.65 \times 10^{25}$ cm$^2$ is the Thomson scattering cross section and 
$N_g \sim 10^{29}$ 
H cm$^{-2}$ is the 
mean gas column density within 0.5 pc (corresponding to a surface density $\sim 10^9 M\odot pc^{-2}$). 
Note that $N_g$ varies by an 
order of 
magnitude above and below the quoted value in the very inner region and in the low density gaps between spiral arms and rings, 
respectively. At the temperature of the core, which is 
$\sim 3000-5000$ K, the opacity due to electron capture by H$^{-}$ might indeed 
be 
up to 10 times higher than that of Thomson scattering for the densities in the inner parsec ($\sim 10^{10}$ g cm${-3}$), making our estimate of the optical 
depth 
conservative. Finally, the gas would then cool on the photon diffusion timescale $t_{diff} \sim H \tau_{es}/c \sim 3000$ yr, where 
$c$ is the speed 
of light and $H \sim  0.1R \sim 0.1$ pc is the vertical scale height of the disk. This timescale is much longer than the orbital 
time, which is $\sim 500$ yr for $R < 1$ pc, meaning that no fragmentation should 
occur. Note also that,as we move out to 2-3 pc the orbital time increases and the optical depth decreases,hence eventually cooling takes
over and fragmentation should happen,which is indeed what we observe in the outer disk (Figure 3,5).

A crucial point is that the diffusion timescale
calculation just performed is independent on metallicity,just depends on density. A higher metallicity would further increase
the true opacity of the gas (namely including all the absorption and scattering processes)  
but by using $\tau_{es}$ we have been already very conservative. The conclusion is thus that our result will hold in general
as long as we can form such a dense central disk. 

\section{The {\it loaded disk} model framework  and its connection with merger-driven inflows}

Building on the results of the hydrodynamical simulations presented in the
previous section, and making use of the concepts of gasdynamics in self-gravitating disks developed
in  section \ref{sec:stability}, in this section we use simple analytical arguments
to sketch a  general scenario for the  physical evolution of an accreting ultra-dense nuclear disk
formed after the major merger of gas-rich galaxies. We will show how the fate of such
disk is ignificantly different from that of a nuclear disk inside an
isolated low-mass protogalaxy at $z \sim 15-20$.

\subsection{Inflows at sub-galactic scales}

As starting point, we consider the dense nuclear disk with the physical properties described in section 3 for the outcome
of the radiative simulations (see 3.3 and 3.4). The characteristic radius of the nuclear disk is then
1-3 pc. While section 3 describes the outcome of simulations carried out in the specific context
of major mergers of massive hi-z galaxies, the size of disks in simulations of protogalaxy formation
in halos of much lower mass ($< 10^{10} M_{\odot}$) at $z > 15$ are also similar (eg Latif \& Volonteri 2015),
as we will recall later in this section. This will allow to generalize our model beyond merger-triggered
nuclear disks. The similarity of radii is only coincidental since in one case the centrifugal barrier
is connected to the mechanism of angular momentum loss in the merger as gas has to flow inward from kpc-scales
in the progenitor galaxies, while in the other case it reflects the fact that the virial radius of high-z
mini-halos is by itself a few hundred pc, which translates then in disks a few pc in size since the centrifugal
radius is of order a few percent of the virial radius (see section 2). We assume the nuclear mini-disk, either
in the merger remnant or in the isolated protogalaxy, to begin in a marginally stable state, $Q \sim 1-1.5$,
simply because statistically this is the most likely state in which a self-gravitating disk would find itself
owing to its tendency towards self-regulation (e.g., Durisen et al. 2007). 
We recall that our simulations described in section 3 confirm that the nuclear disk is in a marginally unstable
state, as shown by the Toomre Q parameter (Figure 7).
In order to simplify the algebra involved in the calculations of this section we will further assume $Q=1$.
In the remainder, we will show how one can determine a critical surface density
$\Sigma_{\rm crit}$, dependent on the gas infall
rate on the disk,  above which fragmentation cannot happen because the cooling time becomes too long. The link
between surface density, infall rate and cooling time will be provided by the optical depth. 

The gas infall rate
can then be related to the expected accretion rate at large scales inside the halo, thus closing the loop
by allowing to relate quantities associated with galaxy assembly in CDM halos to those describing the pc scale dynamics
of nuclear disks.  In the case of an isolated protogalaxy the rationale to achieve that was already explained
in section 2.1, in particular it follows from eq.2. In the case of a major galaxy merger the interpretation of 
eq (2) is that this now describes the inflow down to much smaller scales,
below the centrifugal radius expected at the scale of the galaxy (kiloparsecs), because the latter has been
pushed inward by efficient angular momentum transport triggered by the major merger itself. That is, the centrifugal
radius is much smaller than a few percent of $R_{vir}$, and can only be determined by numerical simulations
such as those of section 3 because it is the product of highly nonlinear dynamics. This is ultimately the pc-scale
characteristic radius that we have already highlighted. We also remark that we are considering specifically major mergers
is important here because it is only in such mergers that angular momentum transport is very efficient (Barnes \& Hernquist
1996; Kazantzidis et al. 2005).
For mergers we can still use the scaling laws between
virial quantities, which now are those of the merger remnant.
For the isolated protogalaxy case a similar approach in spirit was adopted by Lodato \& Natarajan (2006) but the difference
here is the notion that accretion rates at multiple scales play a key role in the evolution of the compact
disk, in particular deciding whether the disk fragments or not. Note that, in the context of protoplanetary
disk formation and evolution, the key role played by accretion has been
highlighted by many authors both in  numerical
simulations and analytical approaches (Boley 2009; Hayfield et al. 2011; Rafikov 2013).
We anticipate that the conclusion will be that the new merger-driven model forms central mini-disks that are much
more resilient to fragmentation compared to disks in conventional protogalaxies. This will
be understood as a consequence of the much higher inflow rates occurring in the merger-driven model which,
in turn, naturally follows from the fact that we consider mergers between the most massive galaxies 
at $z \sim 8-10$ in our model, 2-3 orders of magnitude above the mass of atomic cooling halos considered in the
metal-free, dissociated protogalaxy scenario. In other words, the higher inflow rate just reflects that $\dot M \propto M_{vir}$
in eq(2), where $M_{vir}$ is that of the halo of the merger remnant (which, however, is only a factor of a few different
from that of the progenitor galaxies/halos in a major merger).
Such disks, in turn, can sustain coherent secondary inflows to smaller (sub-pc) scales and form a massive precursor
of the BH seed, whose possible nature will be discussed in the next section.
We will refer to our model as the {\it loaded disk model}. \\

We can start by computing the angular momentum flux for  nearly keplerian
self-gravitating disk  in steady-state (Rafikov 2013;2015) on parsec scales.
Note that, both in the simulations of our merger-driven scenario (MA15),
and in the many published simulations of gas inflows in metal-free protogalaxies (e.g., Latif
\& Volonteri 2015), the
central compact disk that forms has a size of the order of  a parsec. As explained in section 2 in the isolated
protogalaxy case such a size is expected based on rough conservation of angular momentum
of infalling gas into  atomic cooling halos with masses in the range $10^7-10^8 M_{\odot}$
(based on the scaling arguments in eg Mo.Mao \& White 1998). Instead, as we explained above,
in our merger-driven model the pc-scale size reflects efficient loss of angular momentum caused
by the merger dynamics in a much bigger halo, for which the natural size scale would
be of order a kiloparsec had angular momentum been conserved (see, e.g., Mayer et
al. 2007).
These compact disks formed in hydro simulations
are not rotating around a central point-like mass as a bona-fide keplerian disk  but are extremely centrally concentrated since
their early assembly stage. For example, in MA15 we have verified that
the disk rotation velocity profile is
very close to keplerian down to a fraction of a parsec.
The viscous heating originating from  gravitoturbulence, which
is generated by spiral shocks, is assumed to be the main heating source.
This includes also heating from gas-infall as gravitoturbulence will also
be sustained by infalling material keeping the disk self-gravitating.
External heating sources such as UV-radiation produced by the surrounding star forming region
could add to the heating budget hence our calculations in this section should be
regarded as conservative for our purpose, that is to show in which conditions the
disk will remain warm and avoid fragmentation. The disk is then assumed to be {\it self-luminous},
and its state will be determined by the competition between self-generated heat by shocks and
gravitoturbulence on one hand, and radiative cooling on the other hand (Kratter \& Lodato 2016). 
Note that the role of viscous heating in
balancing cooling and opposing fragmentation has been studied and advocated also in 
Latif \& Schleicher (2015), for low-mass protogalaxies at $z > 15$.
We assume the $\alpha$ effective viscosity framework  and, assuming that viscosity is {\it only} provided by gravitoturbulence
in the disk, $\alpha_{SG}=0.1$ (note the subscript to specify that the source is
self-gravity). Indeed $\alpha_{SG}=0.06-0.1$ is the critical threshold for marginally unstable disks.
It corresponds to a maximum {\it local} gravitational stress (see, eg., Lodato \& Rice 2004). 
Following Rafikov (2015) we can then write for the angular momentum flux in a self-gravitating disk subject
to constant mass transport rate $\dot M$:

\begin{equation}
3 \pi c_s^2 \Sigma \alpha_{SG} = \dot M \Omega
\label{eqn:Mdot}
\end{equation}

The keplerian disk assumption allows us to replace the epicyclic frequency 
$\kappa$ with the angular frequency $\Omega$. 
The thermal sound speed $c_s$ will be
fixed by assuming a suitable disk temperature.
Here $\dot M$  is strictly the mass transport {\it through the disk} due to spiral density waves and
gravitoturbulence.
The gas infall from large scales, and its associated infall rate, will be treated as a boundary condition
imposed by the large-scale  properties of the embedding dark matter halo, which determines the potential well. 
In steady-state
the mass transport rate through the disk and the infall rate onto the disk must
match at the disk boundary, hence one can use the latter to set the former.
Indeed, the hydrodynamical simulations presented earlier, show that two rates rates, although time-dependent, are relatively
commensurate just inside and outside the disk. In this case, we can then replace
the $\dot M$ of equation 2 with the typical values of accretion rates found in the
simulation, which are in the range $1000- 5 \times 10^4  M_{\odot}$/yr (see Figure 4). To be conservative, here we assume the
value of $\dot M_{mg} = 1000  M_{\odot}$/yr for the merger case.  For the case of the isolated
disk, we use the value of the typical infall rates measured in the isolated unstable protogalaxy
simulations, ${\dot M}_{iso} \sim 0.1 M_{\odot}$/yr (Latif \& Volonteri 2015).

As mentioned above, in equation \ref{eqn:Mdot} the angular frequency $\Omega$ is replacing
the epicyclic frequency because of the assumption of the disk being keplerian. 
 $\Omega$ is a function of radius, but, for simplicity, we can assume it to be a
constant,  set by the maximum circular velocity of the host dark matter
halo, $V_{max}$, so that $\Omega = V_{max}/r$.  
The  maximum circular velocity
of a NFW halo occurs typically at 2-3 disk scale
lengths, which corresponds to the half-mass radius for a disk
with an exponential mass profile,
and its value depends on the halo virial mass
and the concentration parameter of the halo profile (Bullock et al 2001b).
 $V_{max}$ is typically reached at the halo scale radius $r_s$, which roughly also corresponds to the
disk half mass radius, and also to $\sim 1\%$ of the
virial radius (see, eg., Mo, Mao \& White 1998). $V_{max}$
can be from 10\% to 60\% higher than the
asymptotic circular velocity $V_{circ}$ at the virial radius,
depending on the halo concentration varying in a reasonable range
(5-30, see Figure 1 in Bullock et al. 2001b).
Given that we are now focusing on the region dominated by baryons, we will use
$V_{max}$ in what follows. Halo concentration, in turn, depends
on $M_{vir}$ (eg Figures 4-5 in Bullock et al. 2001b).
Now, for the merger-driven model we can consider the typical  
case of a $10^{12} M_{\odot}$ halo, which for a corresponding mean concentration ($c=12$) 
yields $V_{max} \sim 250$ km/s. Note also that changing the halo concentration by a factor of 2,
which occurs over about two decades in halo mass, changes by only 30\% the ratio $V_{max}/V_{circ}$ (Bullock
et al. 2001b), hence the numbers we are assuming are quite general. 
As the compact mini-disk is located at $< 10$ pc scales, namely at
less than $1\%$ of the virial radius, and the circular velocity profile of massive galaxy
mergers steepens towards the center (see, e.g., Guedes et al 2011), this is a 
lower limit on $V_{max}$ (this will prove to be onservative for our purpose as it will be clear from the
dependencies in the relevant equations below). 
For isolated protogalaxies halo concentration is higher since the halo mass is 3-4 orders of
magnitude lower, hence the concentration will be correspondingly higher.
Yet, as the formation redshift is also higher for mini-halos relative
to the massive galaxies in the merger case, which implies a lower
concentration at fixed halo mass (Bullock et al. 2001b), overall we expect $V_{max}$ to
be again very close to $V_{circ}$.
Solving equation \ref{eqn:Mdot} for $\Sigma$, one can find the {\it characteristic surface density} that 
is consistent with a disk transporting mass via gravitoturbulence at a  rate in equilibrium with
the infall rate from the halo:

\begin{equation}
\Sigma_{eq} = {{\dot M V_{max} } \over {3 \pi {c_s}^2 r \alpha_{SG}}}
\end{equation}

{\it The equation above states the link between large scale and small scale
properties of the inflow.}
We can see that disks with larger mass transport rates embedded in more massive
halos yield proportionally a much higher surface density. This is because
gas can only flow in, namely no outflows are possible as there are no mechanisms
that could generate that. For instance there is no feedback from star formation because
star formation can be neglected. The assumption of no star formation, inspired by
the the results of section 3, will shown to be correct {\it a posteriori}
in order for the model to be self-consistent.
It implies that a more massive central disk has
to assemble from a larger inflow rate. Note that the radius $r$ of the disk here is
a free parameter. It is set by how the specific angular momentum of
gas evolves across multiple scales until fluid elements reach centrifugal support.
As mentioned at the beginning of the section, we will  simply adopt a 
characteristic radius of $1$ pc as we do not model how angular momentum evolves
on scales above that of the nuclear disk. We will explore this aspect further in a future
update of our analytical model. The same characteristic radius can be assumed in the 
isolated protogalaxy case (see Latif \& Volonteri 2015; Latif \& Ferrara 2015 and MA15).
The similarity of the
characteristic radii in the merger and in the isolated protogalaxy case
holds despite orders of magnitude
differences in the virial masses and radii of the systems considered in the two scenarios
because the marked loss of angular momentum in the merger brakes the simple scaling
with $R_{vir}$  for isolated virialized halos (Mo et al. 1998).

From eq. (3) it follows that a  disk accreting at a higher rate will be denser, hence
will be more optically thick, and thus less prone to fragmentation as cooling will be slower.
The higher
the temperature, instead, the lower the surface density, as expected since it
would correspons to a more prominent role of pressure versus self-gravity in
setting the equilibrium in the disk. If we do not
place any strong constraint on radiative properties we can posit that disk can cool efficiently via metal lines
and molecules, down to (at least) a temperature of $1000$ K. This will be used as a reference temperature
below, but we will see that results are not highly sensitive to the assumed temperature.


\subsection{Inflows at sub-galactic scales: results of the {\it loaded disk} model}

We now proceed to turn the previous equations into a fragmentation threshold dependent on mass loading
by exploiting the notion of critical cooling time for fragmentation.
The cooling of the disk, if radiative losses dominate over
other heat transport mechanisms such as convection (see, e.g, Rafikov 2003), is given by the photon diffusion time,
from the disk mid-plane to the surface, which, in line with Section 3, can be expressed by:

\begin{equation}
t_{diff}  \sim {{{H \tau} \over c} \sim  {N_g \sigma_T H \over c}}
\end{equation}

where $N_g$ is the column density, $N_g = \Sigma/m_H$, $m_H$ being the hydrogen mass,
and $H$ is the disk scale height.
We assume the opacity to be given by electron scattering, which is a lower limit on the
total gas opacity, hence will
yields a lower limit on the diffusion time. 
(see also
Mayer et al. 2015). By replacing eq(4) into eq(5) using $N_g = \Sigma_{eq}/m_H$, 
and by setting $\Sigma=\Sigma_{eq}$, we obtain the following  equation for the  diffusion time:

\begin{equation}
t_{diff} \sim \Sigma_{eq}/m_H {\sigma_T H \over c} \sim {\sigma_T \over c m_H}{\dot M V_{max}  H 
\over 3 \pi {c_s}^2 r \alpha_{SG}}
\end{equation}

We recall that we started from the assumption that the disk is in a marginally unstable steady-state
maintained by the balance between cooling, heating and gas accretion, and that as a result it supports
global non-axisymmetric modes that render the flow gravitoturbulent. This state can be realized both
in an isolated protogalaxy and in a merger remnant since the only thing that changes is the way gas accretion
is triggered as well as the value of $\dot M$. It simply means that the characteristic surface
density of the disk will be different in order to obey the steady-state expressed by eq. (4).
Indeed, if we fix $\alpha_{SG} = 0.1$, $r \sim 1$ pc and $H=0.3$ pc (H/r = 0.3 for a thick, gravitoturbulent disk)
, eq(5) can be solved for a given temperature (hence setting the sound
speed $c_s$), a given host halo mass (expressed through the choice of $V_{max}$) and, most importantly, a 
given accretion/inflow
rate $\dot M$. Among the latter three parameters $\dot M$ is the one with the highest impact since,
based on eq. (1), it scales as $M_{vir}$, which can vary by several orders of 
magnitude when going  from the isolated protogalaxy scenario to our merger-driven scenario, while
$V_{max}$ and $c_s$ vary by at most 2  orders of magnitude only  across the range of plausible values of 
temperature and
halo mass (recall $V_{max} \sim M_{vir}^{1/3}$ and $c_s \sim \sqrt T)$. 
Let's now
assume $T=1000$ K, $V_{max} \sim 250$ km/s and ${\dot M}_{mg} = 
1000 M_{\odot}$/yr for the merger driven model, and
$T=1000$ K, $V_{max} = 30$ km/s, ${\dot M}_{iso} = 0.1  M_{\odot}$/yr for the protogalaxy model. 
For the merger-driven model we then obtain that $t_{diff} \sim 10^7$ yr, while for the other model
we obtain $t_{diff} \sim 10^3$ yr. One obtains different orbital times as well, of order $500-1000$ yr and
$> 10^4$ yr, respectively, from $T_{orb} \sim t_{dyn} = 1/\sqrt{G \rho}$ for the average 
central densities  in the two cases, which are, respectively,  $\sim 10^{-12}$ g/cm$^3$ (MA15) and
$\sim 10^{-16}$ g/cm$^3$ (eg Latif \& Volonteri 2015).
The result is thus that for the isolated protogalaxy case the diffusion(cooling) time is shorter than 
the orbital time, making it a strong case for fragmentation. This argues in favour of finding additional
mechanisms to suppress cooling, as it is
customarily done in such a scenario by invoking photodissociation of H2 by UV radiation
and the absence of metals. 
{\it Instead, the diffusion time is comfortably longer than the orbital time because the high accretion rate in the merger case
ensures a high surface density that makes the system optically thick. }
In this case stability to fragmentation is ensured.
A small remark on the temperature, which here is a free parameter, is worthwile. One
would be tempted to relate it to the virial temperature of the halo, and thus to $V_{max}$, in order to eliminate one
variable, but this would be incorrect as the disk temperature is the result of cooling and heating processes
acting at smaller scales. The thermodynamics in the nucleus of the merger remnants
decouples from the simple virial scalings established at the halo level. Moreover, in the merger scenario, if the
disk temperature is set as a boundary condition at the level of the large scale flow, it should not change
owing to the long cooling times. In the simulations of MA15 it is the shock heating of infalling gas that
determines the temperature as a boundary condition (see section 3). The latter temperature is
$10^4$ K, namely higher than our assumed temperature of $1000$ K, but nothing relevant would change in our
quantitative arguments so far due to the weak dependence on $c_s$ in eq. (4,6).

We thus have  shown that, in the simple framework of gravitoturbulent disks in which transport is
simply described by an effective $\alpha$ viscosity, the heavily loaded disk in massive halos is by construction
resilient to fragmentation without the need of any external heating to suppress
radiative cooling. The latter is instead needed in the lighter,mildly loaded central disks 
arising in isolated low mass protogalaxies. Therefore, {\it in our merger-driven model mass loading replaces external
heating sources by maintaining the disk highly self-luminous over long enough timescales,specifically longer 
than the characteristic orbital time}. As anticipated earlier, adding external radiative heating sources, such as from the
starburst in the colder, fragmenting gas just outside the nuclear disk, would only strengthen our conclusions.
It is also important to remark 
that  we have neglected to explicitly include the compressional  heating from the mass loading, which will be more
important in the high inflow rate case of the merger-driven model. Hence our conclusions
are conservative for multiple reasons in this respect. 
Note that our analytical model for the heavily loaded disk in principle extends  the results of the  merger-driven 
scenario to the  general case of high mass infall and transport rate within the disk.
We will discuss in the last section
how this could be relevant to low tidal field, low spin highly biased regions of the cosmic web, which,according to 
recent
cosmological simulations, indeed are preferential sites of very fast accretion rates onto pre-existing BH seeds (Di Matteo
et al. 2017).


The existence of a characteristic surface density for the sustained inflow mode to dominate over
the fragmentation mode is reminiscent of the critical column density
model developed by Larson (2010) to explain the transition between a regime in which
star formation dominates and one in which gas accretion onto a SMBH dominates in
lower redshift systems. The Larson model 
can explain the scaling relation between the mass of SMBHs and the stellar mass of the galactic nucleus or spheroid. In such a model, there exists a gas column density range within which radiative cooling is efficient and gas clouds can become 
Jeans unstable, so that star formation is the the main outcome, while above a critical column density of 
$\sim 1$ g cm$^{3}$ gas clouds become optically thick 
to their own cooling radiation, star formation is suppressed, and gas accretes efficiently onto the central black hole. 
The additional ingredient in our model is the link between optical depth and mass transport rate in the disk, and between that 
and the infall rate from the dark matter halo.

\begin{figure}[]
\includegraphics[width=0.5 \textwidth]{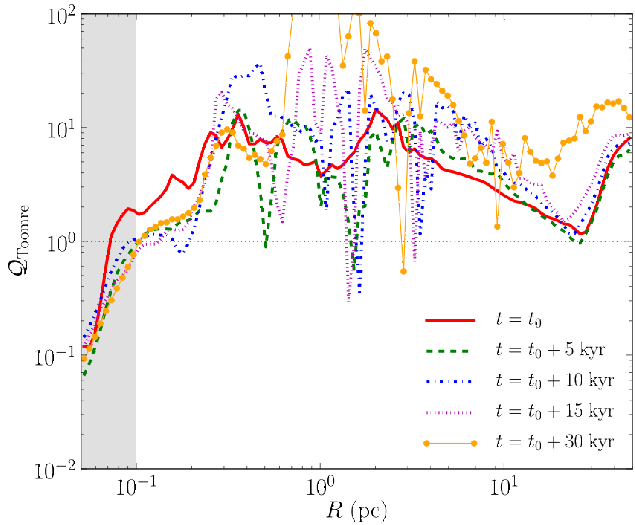}
\caption{Time evolution of the Toomre parameter profile measured from RUN4 (see the adopted definition of the Toomre parameter in section
4.2, which includes the turbulent velocity). 
Dips correspond 
to regions where clumps form from gas fragmentation along spiral arms, predominantly outside the inner parsec region (see Figure 3 for a 
comparison of the spatial distribution of clumps). The dotted horizontal line marks the $Q = 1$ threshold for reference. The gray band 
highlights the  resolution limit given by the gravitational softening. Adapted from MA15.}
\end{figure}

We recall the that the standard $\alpha$ disk is a steady state model, and is only meaningful
for a thin disk in which viscous transport is described in a completely local way
(Lodato \& Rice 2004;2005).
Note that, although it has been often used in the literature, this modelling
framework is clearly very approximate in our case as the circumnuclear disks forming 
in protogalaxies are thick and loaded by accretion of infalling material, and also not
strictly keplerian (although they are centrally concentrated as they form inside-out
from infalling material). Mass transport by spiral modes is global rather than local
in this case since in thick disks  non-axisymmetric modes will have large characteristic 
wavelengths, of order the disk radius,
as clearly seen in many  simulations.
Nevertheless, it has been shown that the global
nature of the transport corresponds, empirically, to an effective $\alpha \sim 1$, namely much larger than the
nominal $0.1$ threshold for fragmentation (see eg Escala et al. 2007). If we had used $\alpha=1$ in the
equations above, differences would be minor owing to the linear dependence in (eq(5)). This
would yield a somewhat shorter diffusion timescale of $\sim 10^6$yr in the merger-driven case, which is still
orders of magnitude longer than the cooling time (note this is very close to the
diffusion time estimate found in MA15 when computing it directly from the properties
of the central mini-disk using eq(4)).
 
Another important caveat is that,although gravitoturbulence is essential in the rationale of the model
just described, its direct effect on the dynamics does not appear in the equations.
This would be correct if
turbulent kinetic energy is rapidly thermalized so that it can be simply absorbed in the choice
of $\alpha$ and $c_s$. 
The turbulent velocity in the MA15 simulations, however, remains higher than the thermal sound speed
at all times usually,which suggests that, at least in the merger-driven  case, it cannot be
simply absorbed in other variables. At the same time, persistent turbulent motions will help
stabilize further the disk against fragmentation (see section 3.2). This could be explicitly treated as a 
{\it viscous heating} term to compare with the diffusion timescale. 
Latif \& Schleicher (2015) have studied the effect of  viscous heating on
protogalactic disks, finding that it has a sizeable effect on  stabilizing the pc-scale disk. However the
effect is smaller than the direct mass loading effect considered here as shown by the fact, in the
case of isolated low-mass protogalaxies,  they
still need some moderate Lyman-Werner flux to avoid fragmentation.

The accretion luminosity can also be added as an extra term but, as shown in MA15, cannot change the basic
features of the physical model presented here in the high inflow rate regimes. A change
of surface density faster than the cooling time might still bring the system to fragmentation adiabatically
(Boley 2009; Hayfield et al. 2011) but if this results from accretion the heating associated should also 
be taken into account, possibly as viscous heating.  These more detailed aspects of the physics of these
loaded  dense disks warrant future investigations.

\section{Final pathways to massive BH seed formation; the notion of {\it dark collapse}} \label{sec:pathways}

The models and simulations described in the previous section provide a scenario for the formation
of a supermassive baryonic disky core that could later collapse into a massive BH seed. However
they do not address how the formation of the BH would actually take place. Indeed several
different scenarios for the final steps leading, or not, to a massive BH seed can be envisioned, and
have been extensively discussed in the literature for the generic direct collapse framework.
In the remainder of this section we will discuss the various pathways using quantitative arguments
directly related to the results of our simulations. We will argue that our merger-driven scenario
lends itself naturally to one particular, intriguing route already proposed in MA15, namely
{\it dark collapse}. In the latter scenario the compact massive gas core of the
mini-disk in the merger remnant could contract further and enter the global General Relativistic (GR)
radial instability regime (Hoyle \& Fowler 1963) without transitioning first to
a SMS or quasi-star stage. In this case a massive black hole would emerge from direct dynamical
collapse of the core, with no intermediate stage.
Perhaps the ony aftermath in this case will be gravitational waves due to the non-zero quadrupole introduced
by deviations from spherical symmetry (e.g,  Saijo \& Hawke 2009), hence the term {\it dark collapse}.

Let us begin with considerations on the final outcome of the MA15 simulations, and in particular on the
structure and stability of the disky core at the resolution limit.  Of course, none of our simulations
includes relativistic dynamics hence the discussion in this section will have to be based on simple
extrapolations and analytical arguments. Also, even close to the resolution limit, despite the very high
densities ($\rho \sim 10^{9}$ g/cm$^3$) and correspondingly ultra-deep gravitational potential well
of the disky core, the system is still well within the newtonian regime until the very end of the
simulations, namely after 50 Kyr from the end of the merger (beyond this timesteps were becoming prohibitively small as
the central density continues to increase, hence we had to stop the calculation).
The first question is what would be the next stage of evolution of the compact core if it 
continues to contract. Would it become a supermassive star (SMS)? With the very high
mass and very high accretion rates occurring in our simulation the SMS is somewhat unlikely.
Schleicher et al. (2013) show that in order to become an SMS the system must reach a threshold
mass dependent on the accretion rate, this being $3.6 \times 10^8 {\dot m}^3 M_{\odot}$. Here $\dot m$ is the accretion
rate in units of solar masses per year. With our rates of $> 10^4 M_{\odot}$/yr
the critical mass is a few orders of magnitude than our core mass of a billion solar masses.
The core could then remain in a massive protoSMS stage for long during collapse as the
heating from Kelvin-Helmoltz contraction will be high enough to control the collapse.

But could the protoSMS enter the GR radial instability phase at some point as it continues
to contract? Let us then consider the conditions
for global relativistic radial collapse. In this case, the mass accretion rate does 
not appear directly in the equations but will still be implicitly instrumental to achieve rapidly
the critical mass for collapse in the first place. Numerical GR simulations show that the instability for a rotating fluid configuration (a polytrope with $\gamma = 4/3$) is reached for (1) a compactness threshold 
$R < 640 GM/c^2$ and (2) a dimensionless spin parameter $q\equiv {cJ}/({{GM}}^{2})\sim 0.97$ (Baumgarte \& Shapiro 1999; Shibata 
\& Shapiro 2002; 
Saijo \& Hawke 2009; Reisswig et al. 2013), where M is the mass and J is the total angular momentum of the cloud, and $c$ is the speed of light.
This has been shown to apply to both uniformly rotating and differentially rotating clouds. The 
inner disky core in our simulations quickly reaches $\sim 10^9 M_{\odot}$ 
within 0.2 pc and it has typically $q \sim 15$. On the other hand, the compactness threshold for $M   \sim
10^9 M_{\odot}$ corresponds to a radius $R \sim 0.03$, which is $\sim 6$ 
times smaller than the characteristic radius of 0.2 pc found
for the final stage of the precursor in the MA15 simulations. Therefore, in the conditions occurring up
to the end of the simulations, the disky core
would be stable to the GR radial instability. 
However,  our simulations also show that the core is on the verge of a global bar instability (see
MA15). Bar modes are barely resolved, hence the instability is artificially suppressed.
which we believe is damped by the fact that the wavelength of the most unstable bar mode is barely
resolved, hence cannot grow. With increased resolution the instability would grow (Pickett et al. 1996).  
Once the core  eventually becomes bar-unstable, transport angular momentum outwards would become very efficient  
and cause further contraction, possibly pushing the system closer to the verge of the instability.

An alternative way to discuss susceptibility to global radial collapse
is to rely on another result of relativistic simulations, which highlights the importance of angular momentum.
Indeed, starting from 2D and 3D rotating polytropic clouds with masses exceeding $10^6 M_{\odot}$, these simulations show that $T_{rot}/W \sim 0.01$ or 
lower is a sufficient condition to bring the cloud to the radial collapse stage under a small initial 
perturbation (Shibata 
\& Shapiro 2002). This is of course a phenomenological criterion, and may change with a different EOS, but 
offers a useful guideline. If we adopt the latter condition, the angular momentum in the inner compact disk has to decrease substantially to 
enter radial collapse since $T_{rot}/W \sim 0.1$ at radii $< 0.5$ pc. 
Drawing from the calculations of bar-unstable protostellar clouds, which 
can  apply here since the eventual subsequent contraction will mostly be in the newtonian regime, one expects a decrease of the 
specific angular momentum, j, 
by a factor of two over a few dynamical times, (e.g., Pickett et al. 1996), i.e., over 
$<10^4$ yr. At the same time, the mass of the system can grow up to a factor of approximately two, if 
accretion rates $>10^4 M_{\odot}/$yr are sustained by the bar down to scales 
$<0.5$ pc for $\sim 10^5$ yr. 
Therefore, $T_{rot}/W$ 
would decrease by almost an order of magnitude at fixed radius on relatively 
short timescales (since it scales as ${T}_{\mathrm{rot}}/W\propto {j}^{2}{M}^{-1}{R}^{-1}$), reaching the critical threshold for radial collapse.

The instability threshold might also be reached first in the inner, denser region. 
Interestingly, we also measured the radial distribution of
angular momentum in the final stage of the disky core evolution in MA15, finding that  $T_{rot}/W$ approaches
the conditions for the GR radial instability progressively more towards the inner region. 
A central small BH seed might thus arise first and continue to accrete from the envelope,
forming essentially a quasi-star (Begelman et al. 2006;2010). However, recent anaytical work 
incorporating the effects of winds and rotation in quasi-stars find that above a mass of
$10^5 M_{\odot}$ there is no steady-state quasi-star solution (Fiacconi \& Rossi 2016;2017).
This basically means that the BH, if formed, will accrete the surrounding envelope on the free-fall timescale
(or on a longer viscous timescale if an accretion disk assembles around it),
effectively reaching quickly $> 10^8 M_{\odot}$ as in the radial instability scenario. 
However, in the latter case the BH seed will shine vigorously as it accretes the entire
envelope dynamically, hence there will be an electromagnetic signal, possibly emitting in the
James Webb Space Telescope (JWST) bands (Volonteri \& Begelman 2010), this being a
major difference from our {\it dark collapse} proposal.

In addition to compactness and residual angular momentum there are, however, additional aspects of the physics that play a role in 
determining the conditions for radial instability.
First of all, the protoSMS that  could result from the disky core may or may not continue to accrete at very high rates.
If high accretion rates continue at small scales, as shown by Ferrara et al. 2013, accretion generates heating that can oppose the 
contraction, and perhaps push the system away from the GR instability regime at any radius . Hence even if collapse continues
it is not guaranteed to end with direct black hole formation.
Second, the detailes of thermodynamics
cease to matter only when the system is already in the GR instability regime, but until then they can
have an impact on the mass growth and in defining the actual threshold for the instability. This means
means that one should at least consider the role of the adiabatic index, even remaining in the simple framework
of a polytropic equation of state.
Numerical GR simulations by Montero et al. 2012, which assume a polytropic protoSMS as an initial configuration,
but turn on nuclear burning from the beginning of their calculation, find that metallicity 
is important below $10^6 M_{\odot}$, potentially causing a phenomenon similar to the pair-instability 
in SN. This can result in deflagration of the SMS before it can collapse into a BH. Of course the latter 
calculations are for non-accreting
objects hence they exaggerate the role of nuclear burning versus heating by accretion and Kelvin-Helmoltz
contraction. Namely, the protoSMS may never become an SMS, as we have argued above following the arguments
of Schleicher et al. (2013).
For non-accreting supermassive polytropes, irrespective on whether the system persists in a protoSMS phase 
or becomes an SMS,  analytical considerations
and direct simulations show that the main physical effect opposing collapse is still rotation.
Including both rotation and a variable adiabatic index for an accreting massive
polytrope, Ferrara et. 2013 are able to derive a critical threshold mass
above which the object should become unstable to the GR radial instability (see
also Latif \& Ferrara 2016). The direct effect of rotation and adiabatic
index is found to be small, and can be encapsulated in scaling factors to determine
the threshold mass, so that one can simply adopt the following expression for the
maximum mass of a stable protoSMS:

\begin{equation}
M_{max} \sim  6 \times 10^5(\dot m/ 1 M_{\odot} yr^{-1}) M_{\odot}
\end{equation}

where the accretion rate is measured is solar masses per year. The above equation can be obtained
if hydrostatic equilibrium holds, namely that the object can continuously adjust to
equilibrium as it grows. The collapse below the threshold mass is stabilized by the heating from
Kelvin-Helmoltz contraction. 

Woods et al. (2017) and Haemmerle et al. (2018) also carried 
out similar calculations for an accreting SMS but using a full stellar evolution code with an 
adaptive nuclear network and post-newtonian
corrections, for accretion rates up to $10 M_{\odot}$/yr. They found that, for the  highest
accretion rate, the critical mass for the relativistic instability is about $3.3 \times 10^5 M_{\odot}$,
which is significantly lower than what is predicted by the above equation. These authors still use the
notion that the SMS can be described as a polytrope and restrict their calculations to
metal-free objects, considering the effect of convection on the
equilibrium state. Interestingly, they conclude that there should be an upper limit for the mass of the
BH seed formed by the radial GR instability, and that this is close to $10^6 M_{\odot}$. The reason is that, 
as the mass accretion rate increases, the SMS collapses earlier, which limits the mass it can gain before
turning into a massive BH. Of course this assumes that there is always an SMS stage preceding the GR
instability, while in our {\it dark collapse} scenario this would not be the case (or, better, the SMS phase
could be so short-lived to be irrelevant for the thermodynamical evolution.)
Revisiting these models
with even higher accretion rates will be interesting, and is currently in progress.
Note that for the extreme accretion rates that we measure
in the MA15 simulations, exceeding $1000 M_{\odot}/$yr, the above equation would indicate
that the system is marginally unstable within the inner pc, which encompassed ($>10^9 M_{\odot}$).
The Woods et al. (2017) results would strengthen further the notion that with such high masses our disky core 
should definitely become GR-unstable, although a more firm result would demand that their calculations
are repeated with our much higher inflow rates.
As Ferrara et al. 2013 point out, for masses above $10^8 M_{\odot}$ there is no simple 
hydrostatic equilibrium solution as the dynamical time becomes shorter than the Kelvin-Helmoltz
contraction timescale. We can add that the analytical treatment brakes down also when the accretion
timescale becomes shorter the KH timescale, because in that case the object cannot readjust to
equilibrium as it gains mass. The whole notion of the polytrope will not be applicable in such conditions.

We conclude that, although a demonstration
will require direct GR simulations of the collapse phase and the ability to model thermodynamics realistically
until collapse ensues, there are strong indications 
that our compact disky core 
may evolve into a radially unstable protoSMS.
Therefore the hypothesis of BH seed formation by {\it dark collapse} appears plausible.
Without a prior SMS phase, up to $\sim 60 - 90\%$ of the progenitor mass can be 
retained during such collapse (e.g., Saijo \& Hawke 2009; Reisswig et al. 2013), the emerging BH seed 
would have a mass roughly between $10^8$ and $10^9 M_{\odot}$, namely in the SMBH mass range 
from the very beginning.
They would be born with masses already close to the mass inferred for the SMBHs powering the high-z QSOs.

If we adopt a more cautious approach, and argue that only a small fraction of the mass of the core
will reach the GR instability regime, such as implied by the results of models with a prior SMS phase, as
in Woods et al. (2017), or involving a quasi-star phase (Volonteri \& Begelman 2010), still even very conservative  assumptions 
highlight that rapid growth
to the required large BH masses to power high-z QSOs is well within reach. To show this, we can follow
the arguments in section 1.2. 
We can assume the target black hole mass is the largest inferred
so far for high-z QSOs,  $M_{BH} = 10^{10} M_{\odot}$, and that $m_{seed} = 2 \times 10^{5} M_{\odot}$, 
corresponding to 0.1\% of the mass of the 
disky core in MA15.
Furthermore, we choose standard parameters $f_{edd} = 1$, or $= 0.1$ and, to compute the characteristic accretion timescale , we adopt a
molecular weight per electron for a plasma at zero metallicity with cosmic abundance of hydrogen (X = 0.75) and helium (Y = 0.25), $\mu_{e} = 1/(1  - Y/2)
= 1.14$, so that  $\tau = 0.395$ Gyr. Note that metallicity effects are marginal in this calculation because they only have a negligible effect
on the value of the molecular weight. With these choices, we obtain $t_(M_{BH}) \sim 0.45$ Gyr. This is compatible with the lookback
time at $z = 6-7$. Assuming that the seed black hole can accrete at the
Eddington limit ($f_{edd} = 1$) is justified by the fact that the hole would accrete the gas belonging to the nuclear disk, which has very high
densities, $n_H  \sim 10^5 - 10^8$ atoms/cm$^3$ at scales below 10 pc.

In summary, there are strong indications that our scenario will lead to a massive BH seed on a fast track, although the actual
mode of formation in the late stages and the mass of the BH seed are yet to be determined.
It remains to be understood if the statistics of the QSO population that should arise, given the properties of the host
galaxies, would match the observations. This will
be the subject of the next section.

\begin{figure}[]
\includegraphics[width=0.98\columnwidth]{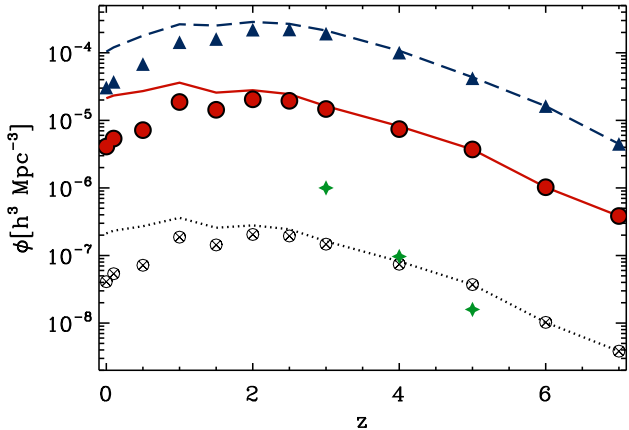}
        \caption{Number density of galaxy major mergers (mass ratio $> 0.3$)   
        hosted by halos with mass above $10^{11} M_{\odot}$. This is obtained by
        multiplying the rate of merger events by a $\Delta t$ (``visibility
        time'') of $100 Myr $ (red solid line). The large red dots indicate the
        number density of events that satisfy the conditions imposed for
        merger-driven direct collapse formation. The blue dashed line and the  
        blue triangles are the same as the red line and symbols, but assuming a
        major merger threshold of $0.1$. The black curve and symbols, instead,
        are the same as the red line and symbols, but assuming a ``visibility time'' of $1
        Myr'$. Finally, the green stars indicate the number density of
        high-redshift quasars calculated by Shen et al. (2007), shown here as
         reference.}
\label{fig:merger_density}
\end{figure}


\section{Merger seed formation in a cosmological context and its observational
consequences: a semi-analytical approach} \label{sec:SAM}

To explore how the scenario of direct collapse formation through galaxy mergers
fits in a cosmological context, in Bonoli et al. 2014 (B14) we generalized the results of MA10 and
conceived a simple analytical model to be incorporated in a framework that follows the
evolution of galaxies in a large cosmological volume. 
In this section, we briefly summarize the results of B14 and discuss future
extensions and applications of the model that will allow to make stronger
predictions on the frequency of merger-driven seed formation and the testability of our
assumptions.  Note that the additional aspects of the formation that emerged
from the simulations of MA15, in particular the possibility of the {\it
dark collapse route}, have not been explored yet in the context of this section,
although a follow-up model of that in B14, which incorporates the new ingredient,
is in progress.

\subsection{A simple generic model for the merger-driven seed formation}

In order study the merger-driven seed model, and its observational consequences,
in a cosmological context, we
generalized the results of MA10 and devised a simple analytical formulation to be
included in a semi-analytical galaxy formation model. 
Semi-analytical
models of galaxy formation are powerful tools to study
 the basic physical processes that drive galaxy evolution, as the global
properties of simulated galaxies can be statistically compared with
observational information (see, e.g., the seminal works of Kauffmann, White \&
Guiderdoni 1993, Cole et al. 1994, Somerville \& Primack 1999). In B14 we made
use of {\it L-galaxies}, the
semi-analytic model from the Munich group, in the version presented in Croton et
al. 2006 and De Lucia \& Blaizot 2007, and run on the outputs of the N-body Millennium
simulation (Springel et al. 2005). With this approach, we were able to follow the
cosmological evolution of galaxies residing in halos above $\sim 10^{10} h^{-1}
M_{\odot}$, limit set by the mass resolution of the Millennium simulation. In
what follows, we explain some details of the model and the main results obtained
in B14.

 In every new dark matter halo identified in the simulation, a new
galaxy is initiated with the baryon fraction corresponding to the assumed
cosmological model. After being initalized, galaxies start evolving through gas
cooling, star formation, feedback, and other physical processes as explained
in  Croton et
al. 2006 and De Lucia \& Blaizot 2007. In B14 we assume that, when a galaxy is created,  a first generation
of stars had already been able to create a seed black hole from a PopIII
remnant. As we follow the hierarchical evolution of each galaxy, we assume that
a galaxy merger event can either lead to the growth of the PopIII remnants via merger
and gas accretion, or can lead to the creation of a merger-driven massive seed,
if the merger event approximates the initial conditions of the
simulations of M10. Specifically, we require  (i) that a  major merger has taken
place (in our default model, we set as threshold for major merger a mass ratio between the
stellar and gas components of the two galaxies of $1:3$), (ii) that the
merger remnant is hosted by a halo of at least $10^{11} m_{\odot}$,
 (iii) that the merging galaxies are heavily disk-dominated, with   
bulge-to-total ratios of, at most, $0.2$ and, finally, (iv) the
  lack of a pre-existing black hole above $10^6 M_{\odot}$. This last condition
is based on energetic arguments, as a pre-existing black hole of already
intermediate mass would likely be able, though feedback, to eventually stop the gas inflow needed
to form the massive BH precursor.   This, of course, would hold in the simple
case of isotropic heating; if, instead, radiation is anisotropic,
it could still be possible for gas to flow-in at high rates and perhaps lead to  Super-Eddington
 accretion phases  on the pre-existing black hole (see Section
\ref{sec:conclusions} for a
discussion on Super-Eddington accretion).

 When the  requirements listed above are met, we assume that a massive seed is able to form
at the center of the merger remnant, after a rapid quasi-star, SMS or protoSMS phase
(see Section \ref{sec:pathways}).

In Figure \ref{fig:merger_density} we show the number density of mergers that satisfy the above
conditions, and thus lead to the formation of a massive seed.
Note that from the merger rate, which is the information provided by the
simulation, we obtain a value for the number density of  mergers  by assuming a merger
``visibility'' time of $100$ Myr. With this assumption, the number density of
all galaxy major mergers, hosted by halos above $10^{11} M_{\odot}$,
is shown by the red solid line. The red bullets show instead the number density
of the mergers that can  lead to the formation of a massive seed, that is, the
mergers that satisfy all the other conditions imposed. We see that at
high redshift, most mergers could lead to a merger-driven seed.
To guide the eye with an order-of-magnitude comparison, we plotted in the same
figure the observed number density of optical quasars reported by
Shen et al. 2007.
While a direct comparison between the observed quasar properties and
our black hole population is beyond the scope of this paper, we see that the
number of
events possibly forming direct collapse seeds is large enough to account for the 
bright optical quasars. This would hold even assuming a much shorter visibility
time of $1$ Myr (dotted black curve and black symbols). 
In the same Figure, the blue dashed line and the blue triangles show the number
density of mergers and events of massive seed formation if the threshold for a
major merger is decreased from $1:3$ to $1:10$ (as used in
 Volonteri \& Begelman 2010
in their study  of  the
 evolution of quasi-stars). Clearly, the number of events assuming a smaller
 mass-ratio threshold is much higher, approximately an order of magnitude higher
 at all redshifts.

\begin{figure}[]
\includegraphics[width=0.98\columnwidth]{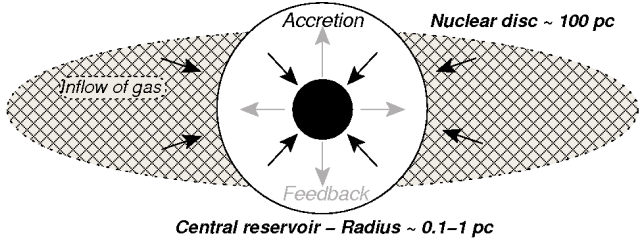} 
\caption{Sketch of the structure around a newly-formed massive seed after a   
        major merges. The seed is surrounded by a gas ``reservoir'' (white area), which is
        composed of the remnants of the central massive cloud from which the    
        black hole formed in the first place, and from gas still flowing from
        the nuclear disk (squared gray region). The growth of the seed stops   
        once its feedback energy balances the binding energy of the central
        reservoir.}
\label{fig:sketch}
\end{figure}

After formation, the massive seed  is assumed to be
able to  start accreating at the Eddington rate any gas still available in a gas
``reservoir'', generated from any residual gas inflow from the circumnuclear
disk. The seed stops accreting from this gas reservoir once its feedback is able
to unbind the reservoir itself (see Figure \ref{fig:sketch} for a sketch of the structure
surrounding the massive seed soon after formation). In this simple picture, we
assumed isotropic and thermal feedback. In the B14 default model, the feedback
efficiency is assumed to be $5\%$ and the central reservoir to have a radius
of $1$pc.

\begin{figure}[]
\includegraphics[width=1.\columnwidth]{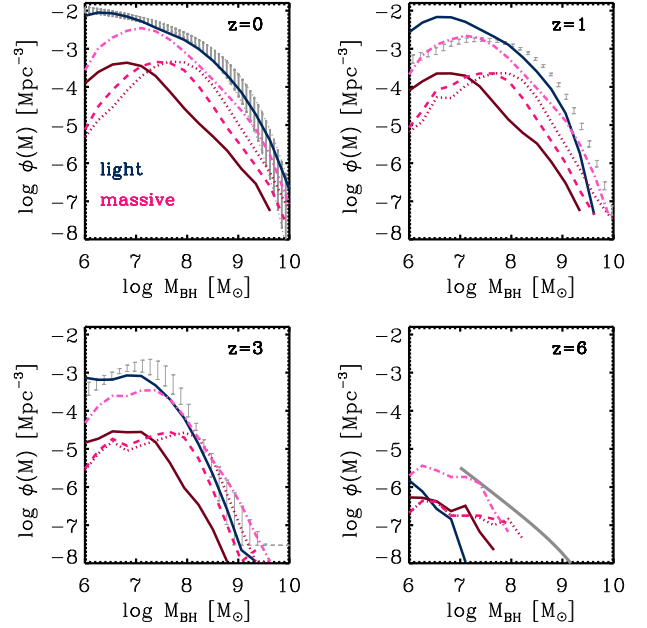}
        \caption{Black hole mass function predicted by the model for the light
        seed remnants (blue solid curve) and for the merger-driven seeds (red
        lines),
        assuming different flavours of the
        model. The solid line refers to the reference model (reservoir of
        $1$pc size, $5\%$ efficiency of feedback coupling, and minimum mass
        ratio for major mergers of $1:3$). Varying one parameter at a time, for the dotted
        line we assumed a reservoir of $0.1$ pc , for the dashed line a feedback
        efficiency of $1 \%$, and for dot-dashed line a minimum mass ratio of
        $1:10$ for major mergers. The gray curves show observational estimates
        of the black hole mass function (Shankar et al. 2004 at z=0, Merloni \&
        Heinz 2008 at z=1 and z=3, and Willott et al. 2010b at z=6)}
\label{fig:BHMF}
\end{figure}

In Fig. \ref{fig:BHMF} we show, at various redshifts, the resulting black hole
mass function, separated depending on the 
  progenitor black hole: the blue
solid curve shows the mass function of black holes with a PopIII progenitor,
while the different red curves show the mass function of the descendants of
merger-driven massive seeds, for different flavours of the model. As mentioned
earlier, in our
reference merger-seeding model (solid red line), we have assumed a baryonic mass ratio of
$1:3$ for the major merger threshold, a size of $1 $ pc for the gas ``reservoir`` from which
the newly formed black holes can accrete, and an accretion feedback efficiency
of the growing massive seed of $5\%$. The different red curves in the figure
show how sensitive our results are on the choice of these parameters.
Compared to the reference model, we get a shift of the mass function to higher
masses if we either decrease the physical size of the reservoir (dotted red line) or decrease the
efficiency of feedback (dashed red lines) as, in both
cases, the newly-formed seed can grow to higher masses before its feedback 
energy is able to
unbind the surrounding gas reservoir. When we instead
lower the minimum mass ratio for defining a major merger (dotted-dashed red
lines), we see an increase of
the normalization of the mass function at all masses, as the number of events
that lead to the formation of a massive seed increases by at least an order of 
magnitude with respect to the reference model.
While light seeds descendants are the dominant population at low-redshift, the
fraction of black holes with a massive progenitor increases with increasing
redshift and mass.
Also, we note that at the highest redshift
shown in the figure,  to reach the masses powering the most
luminous quasars (gray line, Willott et al. 2010b), it seems that  more frequent and more massive direct
collapse seeds are  necessary. We note, however, that the volume probed by the
Millennium simulation is not large enough for an exhaustive statistical study
 of the rare high-z quasars. As further discussed below, we are preparing a new
study  which include a more comprehensive model for BH seeds and uses the
Millennium-XXL simulation (Angulo et al. 2012), which simulates a cosmological volume $216$ times
larger.

\begin{figure}[]
\includegraphics[width=1.\columnwidth]{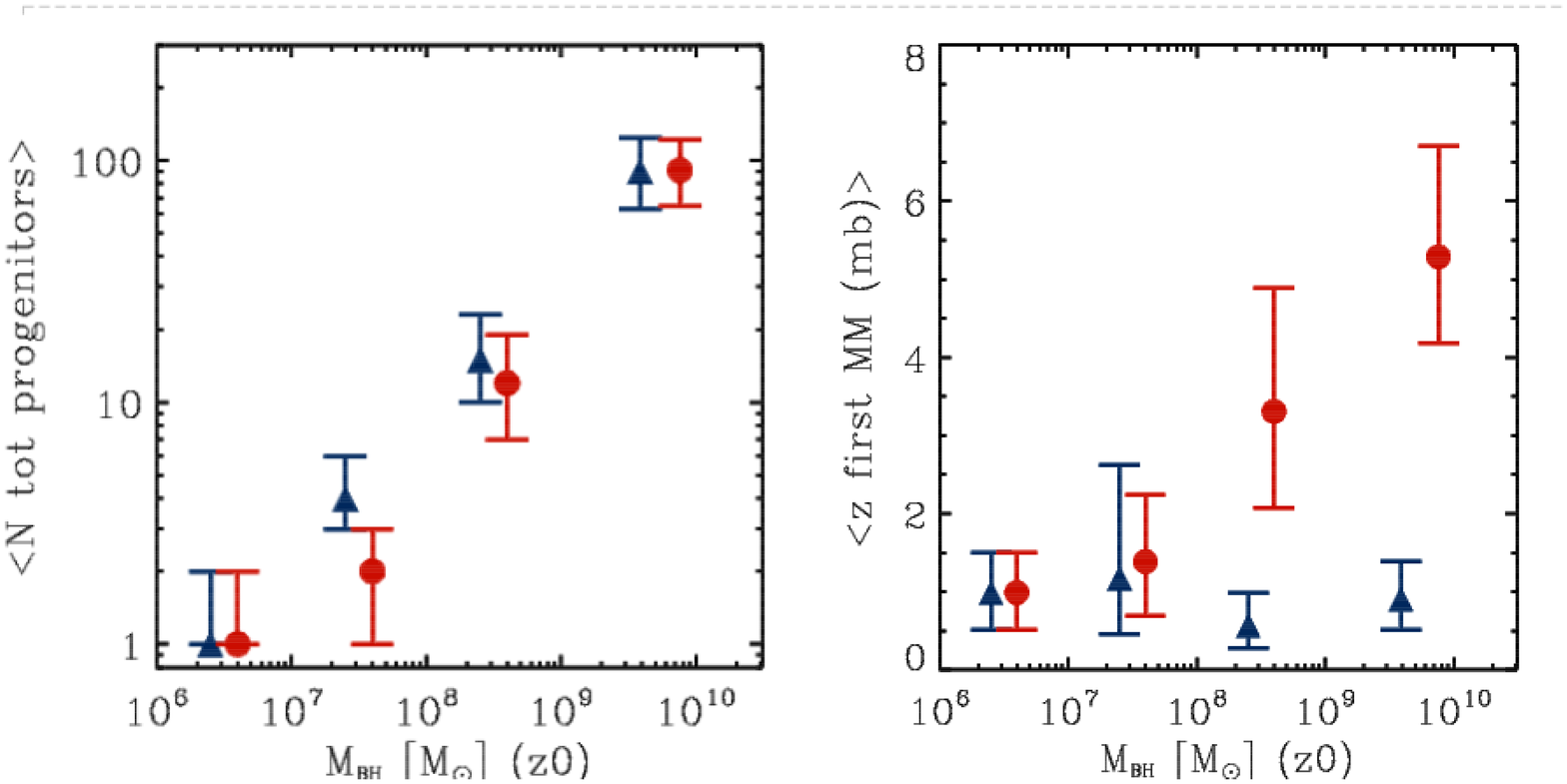}
        \caption{Merger histories of the galaxies hosting light-seeds
        descendants (blue triangles) and merger-driven seeds (red circles). As a
        function of final black hole mass, we show, on the left, the median
        number of progenitors and, on the right, the median redshift of the
        first major merger. The error bars bracket the 25th and 75th percentiles. }
\label{fig:history}
\end{figure}

\begin{figure}[]
\includegraphics[width=1.\columnwidth]{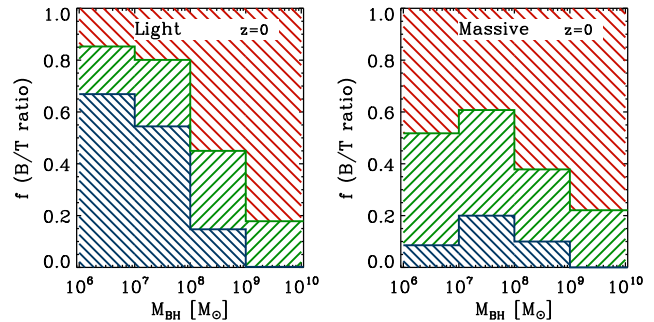}
\caption{Morphology of BH hosts. Left panel: at fixed final black hole mass,
        distribution of the morphology (B/T ratio) of the galaxies hosting light
        seeds descendants. Blue refers to disks or extreme late-types
        (B/T$<0.3$), green refers to normal spirals ($0.3<$B/T$<0.7$), while red
        refers to ellipticals (B/T$>0.7$). Right panel: the same, but for massive seed
        descendants. While the most massive black holes are hosted primarily by
        early type galaxies, independent of the progenitor, smaller black holes
        descendant of massive seeds are generally in more bulge-dominated
        galaxies than the same black holes descendant of light seeds.}
\label{fig:hosts}
\end{figure}

Finally, one of the advantages of studying a merger-driven BH seed scenario
within the framework of  a galaxy
formation semi-analytical model, is that not only we can make predictions for
the black hole population, but we can also study the
properties of the galaxies that host them and their environment. For example, when comparing the
merger histories of galaxies that host black holes descendants of massive seeds
and of light seeds, we find interesting differences (see Fig. \ref{fig:history}): at
fixed final black hole mass, while the total number of progenitor galaxies is similar
for the light and massive seeds remnants, the typical redshift of the first major merger
is very different for the two cases. Galaxies that experienced a major merger early on
in their history are much more likely to host a merger-driven seed. This simple
result can have important implication in the type of environment in which light
and massive seed descendants can be found. Moreover, we also find that, at fixed final black hole mass,
massive seeds are more likely hosted by early-type galaxies than light
seeds (see Fig. \ref{fig:hosts}).
Also, while our reference assumptions for the parameters determining the formation and
first growth episode of massive seeds lead to black holes that are consistent
with the MBH-MBulge relation, some variations of our reference model, such as a
smaller size for the  gas reservoir around the new massive seed, lead to a
population of black hole that lie above the MBH-MBulge local relation. Such model predictions can be
directly tested once a full-census of  local 
black holes is available.

\subsection{A broader look to the black hole seed population}
 
As discussed above, in B14 we studied the frequency of merger-driven seed
formation
by incorporating a simple model for the creation of seeds into a galaxy
formation framework that tracks the evolution of galaxies in a cosmological context.
With this approach, we have been able to study not only the statistics of formation
of merger-seeds and properties of their descendants, but also the environment in which
 these seeds evolve down to the present time.
 Following-up the results of MA15, an update of B14 is necessary to revise the
 occurrence of the specific conditions that can lead to a massive
seed. In MA15 we briefly investigated the statistics of  $z>6$ major mergers  in halos
above $10^{12} M_{\odot}$, which could potentially sustain strong enough central
inflow rates causing a ``cold direct collapse'' and the direct
formation of  $\sim 10^8 M_{\odot}$ seeds. Using the Millennium simulation, we
calculated that, by $z\sim 6$, there are  $\sim 3 \times 10^{-8} \rm{Mpc}^{-3}$
galaxies that live in halos above $10^{12} M_{\odot}$ and already experienced a major
merger. This  number density is comparable to the number density of
known luminous high-z quasars, which is $\sim 10^{-8} \rm{Mpc}^{-3}$ (Willott et
al. 2010).
Work is thus in progress to  refine the  prescriptions adopted in B14 for the
 modelling of different seed scenarios, and study in more details the
 environment of the descendants of the different seed populations. In this work
 in progress, we are exploiting simultaneously the simulations MillenniumII
 (Boylan-Kolchin et al. 2009),
 Millennium (Springel et al. 2005) and Millennium XXL (Angulo et al. 2012),
 which, combined, offer an extremely large mass
 dynamic range.
Guo et al. 2011, for example, already used the combination of the Millennium and MillenniumII simulations to study
the evolution of a wide range of galaxy populations, from dwarfs to large
ellipticals.
 For our purposes, such approach opens the possibility of exploring, at once, the evolution of
 PopIII seeds, likely developing in more common environments, and of direct
collapse generated in 
 the more rare high-density peaks.

\section{Alternative scenarios for the origin of bright high-z QSOs and
observational prospects} \label{sec:conclusions}

In this article we have focused on the description of a new scenario for the
formation of massive BH seeds, highlighting the main physical difference with
respect to the more standard direct collapse models. 

A very different, alternative perspective on the problem of the origin of the
extremely massive black holes powering high-z quasars,  relies
on the notion that the growth rate of initially light seeds, such as those from PopIII stellar remnants,
 could be much higher than the ones established by the conventional Eddington
limit (Madau et al. 2014; Sadowski et al. 2015). Interestingly, the powerful
inflows triggered by catastrophic angular momentum loss which are at the base of
our merger-driven direct collapse scenario, can be instrumental to such alternative approach. Indeed, if a light seed could
grow at a rate comparable to the inflow rates found in our simulations as well as in our analytical model
of loaded disks, clearly the BH light seed could reach the same mass of our supermassive cloud precursor in
 similarly short timescales. As discussed in the previous sections, if there is
a pre-existing black hole with mass $10^6 M_{\odot}$ shining at near Eddington luminosity,
rapid gas infall could be stifled by radiative feedback. However, if the seed is
a few hundred solar masses, its initial accretion luminosity would be very low, and it could grow quickly to
at least $10^6 M_{\odot}$ based on the same  argument. Further growth could
continue  if the radiative efficiency
is very low, as  in typical Super-Eddington accretion models (e.g., Madau et al. 2014),
or if feedback is somehow ineffective at producing enough radiation pressure to overcome the
infall of matter. We refer to Johnson \& Haardt (2015) for a recent review on
the subject of radiative feedback. Here we just 
recall the essential aspects of Super-Eddington accretion and the state of
research in this topic.

Models of
radiatively inefficient accretion discs, such as the SLIM models, in which
radiated photons are advected to the black hole instead
of diffusing on the photon transport timescale in optically thick gas,  provide a conceptual framework
 for super-Eddington accretion (Sadowski et al. 2015). Equipped with a SLIM disk solution, semi-analytical models
of black hole growth have shown that a light seed of a few hundred solar masses could grow to several
billion solar masses in less than a billion years (Madau et al. 2014). In these
models, rapid growth essentially
arises because of a very low radiative efficiency $\epsilon _{r} < 0.01$ which
boosts mass growth  (see equation Eqn. \ref{eqn:tBH}).  There are however two
potential issues.
First of all, the SLIM disk solution is a simple-steady state solution which does not investigate directly the coupling
between radiation and advection via turbulence, despite angular momentum transport in accretion disks is believed
to be associated with turbulence triggered by processes such as the magnetorotational instability.
State-of-the-art radiation 3D hydrodynamics simulations, such as those
by Jiang et al. (2014), which adopt sophisticated radiation transport schemes that naturally allow for anisotropic
radiation propagation, find results that differ from the SLIM disk solution. In
such simulations,  Super-Eddington
flows do arise but they are at least an order of magnitude more radiatively efficient than in the SLIM disk solution. 
However, these simulations have been  carried out only for stellar mass black holes 
and for isolated accretion disks, which have very different flow conditions and
environment relative to  Pop III seed remnants in highly  accreting mini-halos.

Recent 2D simulations for BH seeds of $> 10^3 M_{\odot}$, at the high
mass end of the Pop III remnants mass distribution (Hosokawa et al. 2013; Hirano et al. 2014) in high density fluid disks,
hence more appropriate for the high-z QSOs application, have been published
(Inayoshi et al. 2015; Takeo et al. 2018).
They have confirmed the existence of Super-Eddington accretion in the equatorial plane while radiation leaks out
mostly from the poles.  For example, Takeo et al. 2018 find that
extremely high accretion rates are possible, in the range 10-100 times higher than the Eddington rate,
yet their results do not stem from a fully-self consistent solution of the
radiation hydro equations, rather they specify an
anisotropic radiation flux a priori, and then apply the diffusion equation within the propagation cones.
These flows are of the order of 50\% of the Bondi rate, thus
are essentially limited by the larger scale supply rate.
However, simulations of mini-halos hosting Pop III
remnants show that the gas is warm, diffuse and ionized in such 
environments, with densities of $1-10$ cm$^{-3}$ (Johnson
\& Bromm 2006) rather than the $\sim 10^5$ atoms/cm$^3$ assumed in Takeo et al.
2018. This means the Bondi
accretion rate, which is proportional to density,
 is highly overestimated in Takeo et al. (2018) for the typical case of a BH seed
produced by the collapse of a Pop II star.
Instead, the conditions for rapid growth via Super-Eddington accretion might be more favourable later on, after the ambient gas
has cooled and recombined, raising the characteristic density.
This is plausible at $z \sim 10$, and even more so in more massive halos in which
gas has collapsed to higher densities to form the central galaxy. These of  course would be atomic cooling halos,
with virial masses $> 10^8 M_{\odot}$.

Lupi et al. (2016) explored 
such "late Super-Eddington growth" scenario,  addressing  it in 3D adaptive mesh refinement (AMR) and
lagrangian mesh-less hydrodynamical simulations of
circumunclear disks hosted by massive high-z galaxies. By parametrizing the
growth of several embedded BHs using an effective (low) radiative
efficiency from the SLIM disk model, they do find rapid growth of a light BH
seed ($\sim 100 M_{\odot}$) to a million solar, in $< 10^7$ yr,
but cannot follow that further due to the limited volume of the simulations which do not include gas accretion. In such
conditions, indeed, growth becomes limited by the gas supply.
Moreover, they found that  isotropic radiative heating by massive BHs accreting in a spherically symmetric
fashion, a popular form of ``AGN feedback''  adopted in simulations of  black hole
growth (e.g., Di Matteo et a. 2012; 
Feng et al. 2014), 
cannot stifle Super-Eddington accretion. Whether or not circumnuclear disks are common in the nuclei of high-redshift galaxies
is yet to be seen. A cold self-gravitating disk is instrumental to generate prominent gas inflows via spiral instabilities,
which is qualitatively similar to what happens in the merger-driven model. High
inflow rates, of the order of the Bondi rate, are also found in simulations
of circumnuclear disks (Souza-Lima et al. 2016). 
What remains to be seen is if such compact disks actually exist in high-z
galaxies. In galaxies of typical size and mass at $z \sim 7-8$,
this appears unlikely, as stellar and SN feedback drive hot outflows and help to generate a multi-phase, turbulent thick disk
(Escala et al 2007; Fiacconi et al. 2017) in which the dynamics is very
different from that in Lupi et al. (2016). In such systems  
accretion is  going to be highly 
anisotropic as it is driven mostly by accreting cosmological filaments (Fiacconi et al. 2017). Locally, the inflow rate could still become
Super-Eddington, hence accretion rates could be highly fluctuating. Sustained, episodic Super-Eddington accretion has been proposed
as the mode by which even relatively light seeds can grow to very large masses despite feedback effects,
possibly explaining also why the fraction of active massive BHs is small at any time despite overall significant
growth (Pezzulli et al. 2017a).

It thus appears that growth via Super-Eddington accretion might be more 
plausible in more massive halos which
arise towards $z \sim 10$. Halos as massive as those considered in our merger-driven model at $z \sim 10$ would certainly
belong to this category. Mergers would not be necessarily involved but they could aid growth by bringing a large
reservoir of gas to parsec scales as the sphere of influence of light Pop III seeds would be quite small.
Therefore, the conditions favourable to our scenario would also be very favourable to Super-Eddington accretion.
While feedback by the growing black hole would be a difference in the two cases, Lupi et al. (2016) suggests that 
growth rates are not suppressed
because the central black hole seed can augment its mass rapidly  by engulfing an entire cluster of stellar and intermediate 
mass black holes  rather than just via gas accretion.
New work in progress  uses hydrodynamical simulations of massive galaxy formation at $z \sim 10$ with unprecedented 
resolution, based on re-sampling to sub-pc scales the  {\it MassiveBlack} runs
used in Di Matteo et al. (2012),
to study the growth of high-z quasars  from already sizeable seeds. Preliminary results suggest
that, even before a major merger, tidally interacting massive galaxies at $z > 6$ can form a massive ultra-compact
disk  due to efficient removal of angular momentum by tidal perturbations (Capelo et al. in prep.). This could 
either give a head-start to supermassive cloud precursors
of direct collapse black holes, or fuel Super-Eddington accretion. Since resetting the local angular momentum 
to low values by whatever mechanism seems the key to obtain ultra-dense central baryonic cores, the 
local tidal field in the cosmic web
might play a significant role in setting the properties of the highly biased
hosts of quasars at high-z. Recent results from
the {\it Blue Tide} simulation show indeed a tendency for rapid growth of
massive BH seeds to be favoured in regions with a low
tidal field, and correspondingly galaxies appear more centrally concentrated in
such regions (Di Matteo et al. 2017). Sustained
multi-scale inflows in such galaxies could be triggered without the need of mergers due to the fact that angular momentum
is much lower to begin with. The preliminary results of Capelo et al. (in prep.)
also show that a well defined disk is present since $z \sim 10$ in such galaxies, 
although
it is thick and has a relatively high velocity dispersion. Such disks are considerably more gas-rich (with $50\%$ gas
fractions) than the models employed in MA10 and MA15, and globally unstable to large amplitude spiral modes and bar-like 
modes. Although future analysis will have to address gasdynamics to sub-pc
scales, these results  suggest that the thin, non-turbulent, 
stable disk models adopted in the initial conditions of  MA10 and  MA15 were on the conservative side. Namely, with
realistic high-z disks, angular momentum loss could be significant even before the merger is completed.

Observationally, determining the census of very high-z bright quasars at even
higher redshift than the ones known today (e.g., in the redshift 
range $8-15$), might be
the only way to disentangle between merger-driven direct collapse and
Super-Eddington growth of light seeds. The latter regime has
to wait to $z < 10$, after gas previously ionized in mini-halos has fully
recombined, for cooling to happen and gas to be able to feed
an eventual BH seed again. Instead, in our merger-driven model, especially in its {\it dark collapse} variant, seeds can be
so massive from the very start that they could reach $> 10^8 M_{\odot}$ already
at $z > 10$, thus powering quasars with
$L_X \sim 10^{43}$ erg/s already then. Also, if such population exists already at $z = 8-10$, but not earlier, it might
favour our scenario against the more conventional scenario of $H_2$-dissociated
protogalaxies, as it would be consistent
with the notion that sufficiently massive, metal-enriched  halos have to appear
first for direct collapse to be triggered.
The ATHENA X-ray observatory should detect hundreds of AGNs at such very high redshifts, providing very powerful new 
constraints on competing BH seed formation and/or growth scenarios (Aird et al. 2013).
 
In addition, studying the evolution of the quasar luminosity function in the redshift range
$z \sim 5-15$ will be crucial to test multiple scenarios. Indeed, it is likely that black holes can form and grow via two
or more of the channels proposed so far, which hopefully can be distinguished
through the different efficiency at which  black holes
 grow. Indeed, if, as it seems currently, there is a dearth of lower luminosity
AGNs in the $z \sim 5-10$ range (Treister et al. 2013;
Weigel et al. 2015; Pezzulli et al. 2017b), it may imply that lower mass black
holes have yet to grow, perhaps from more common, lighter seeds, and that
Super-Eddington accretion is not a common route. At the same time, it would
imply that the population of bright high-z quasars is
special, and likely produced by a direct collapse channel.
LISA will provide the only direct way to probe the particular {\it dark collapse} idea, since ultra-massive seeds forming in this way will emit
a gravitational wave signal during their asymmetric collapse, which can be calculated and is expected to fall in the low frequency
band and above the sensitivity curve of the current instrument design (e.g.,  Saijo \& Hawke 2009).

There have been recent claims that direct collapse black holes have been
observed, in particular for CR7, an estimated 
  bright Lyman alpha  emitter at $z = 6.6$ for which no metal lines could be detected (Sobral et al. 2016; Pacucci et al. 2016). While the presence of a 
massive black  hole powering
the emitter is a highly plausible hypothesis (Smith et al. 2016), the absence of metal lines in the spectrum, which would be 
consistent with the conventional notion that direct collapse black holes require
metal-free gas, is not
constraining the formation process, only the nature of the accreting gas. Therefore, a light seed black hole
from Pop III stars, that has subsequently grown by Super-Eddington accretion in relatively pristine gas, is an 
equally plausible hypothesis. Detecting a similar source at $z > 15$ would instead become interesting in that
it would start to really favour direct collapse in metal-free gas as Pop III remnants might still be embedded
in ionized, low-density gas at those early epochs (see Section 1), in which Super-Eddington accretion is
not feasible.
Probing the nuclear cold gas distribution of high -z massive galaxies, in a temperature range below $10^4$ K,
as in our compact nuclear disk, could provide also an indirect way to test model assumptions, especially 
to verify if the extremely  high densities and supersonic infall velocities
arising in our merger scenario did take place in some systems. 
The combination of powerful instruments probing the very high redshift Universe at all wavelengths, from ALMA
to JWST, ATHENA, WFIRST and SKA, holds the promise to shed light on the fascinating subject of massive black hole formation.

\acknowledgements
The authors are thankful to the Kavli Institute for Theoretical Physics (KITP) at UC Santa Barbara for hospitality during April-June 2017
during the Program "Galaxy-Halo Connection" where the loaded disk model framework was originally conceived and elaborated. They also acknowledge
useful and stimulating discussions on merger-driven direct collapse through the years  with M. Begelman, A. Ferrara, M. Latif, R. Schneider, Z. Haiman, 
P. Natarajan, N. Yoshida, M. Colpi, D. Fiacconi, E. Rossi, K. Schawinski, T. Di Matteo, M. Davies, M. Volonteri, P. Capelo,  P. Montero, L. Rezzolla, D. Schleicher.


\begin{thebibliography}{}



\bibitem[Agertz et al.(2009)]{agertz+09} Agertz, O., Teyssier, R., \& Moore, B.\ 2009, MNRAS, 397, L64

\bibitem[A (2009)]{A2008}Abel, T., Bryan, G. L. \& Norman, M.L., 2002, Science, 295, 93
\bibitem[A (2009)]{A2008}Ade et al.,  Plack collaboration \ 2016, A\&A, 594, 13
\bibitem[Agertz et al.(2009)]{agertz+09} Agertz, O., Teyssier, R., \& Moore, B.\ 2009, MNRAS, 397, L64
\bibitem[A (2009)]{A2008}Agarwal, B., et al 2012
\bibitem[Alvarez et al.(2009)]{alvarez+09} Alvarez, M.~A., Wise, J.~H., \& Abel, T.\ 2009, \apjl, 701, L133
\bibitem[Amaro-Seoane et al.(2013)]{amaro-seoane+13} Amaro-Seoane, P., Aoudia, S., Babak, S., et al.\ 2013, GW Notes, Vol.~6, p.~4-110, 6, 4
\bibitem[Angulo et al. (2012)]{angulo+12} Angulo, R., et al.\ 2012, \mnras, 426, 2046
\bibitem[A (2009)]{A2008} Ba\~ nados, E., et al.  2018, Nature , 533, 473
\bibitem[A (2009)]{A2008} Barnes, J.E, \& Hernquist, L. , 1996, ApJ, 471, 115
\bibitem[Baumgarte \& Shapiro(1999)]{baumgarte+99} Baumgarte, T.~W., \& Shapiro, S.~L.\ 1999, \apj, 526, 941
\bibitem[Begelman et al.(2006)]{begelman+06} Begelman, M.~C., Volonteri, M., \& Rees, M.~J.\ 2006, \mnras, 370, 289
\bibitem[Begelman et al.(2008)]{begelman+08} Begelman, M.~C., Rossi, E.~M., \& Armitage, P.~J.\ 2008, \mnras, 387, 1649
\bibitem[Begelman \& Shlosman(2009)]{begelman+09} Begelman, M.~C., \& Shlosman, I.\ 2009, \apjl, 702, L5
\bibitem[Begelman(2010)]{begelman+10} Begelman, M.~C.\ 2010, \mnras, 402, 673
\bibitem[Begelman(2012)]{begelman+12} Begelman, M.~C.\ 2012, \apjl, 749, L3
\bibitem[A (2009)]{A2008} Behroozi, P. , Wechsler, R.H., \& Conroy, C., 2013, ApJ, 770, 57
\bibitem[Belkus et al.(2007)]{belkus+07} Belkus, H., Van Bever, J., \& Vanbeveren, D.\ 2007, \apj, 659, 1576
\bibitem[Boley(2009)]{boley+09} Boley, A.~C.\ 2009, \apjl, 695, L53
\bibitem[Boley et al.(2010)]{boley+10} Boley, A.~C., Hayfield, T., Mayer, L., \& Durisen, R.~H.\ 2010, \icarus, 207, 509
\bibitem[Boylan-Kolchin et al. (2009)]{Boylankolchin+09} Boylan-Kolchin, M. et al. \ 2009, \mnras, 398, 1150
\bibitem[A (2009)]{A2008} Bromm, V. \& Loeb, A., 2003, ApJ, 596, 34
\bibitem[A (2009)]{A2008} Bromm, V., \& Larson, R.B., 2004, ARA\&A, 42, 79
\bibitem[Bonoli et al.(2014)]{bonoli+14} Bonoli, S., Mayer, L., \& Callegari, S.\ 2014, \mnras, 437, 1576
\bibitem[A (2009)]{A2008} Brandt, W.N. \& Hasinger, G. \ 2005, ARA\&A, 43, 827
\bibitem[A (2009)]{A2008} Bullock, J. S., et al.,	2001 ApJ, 555, 240
\bibitem[A (2009)]{A2008} Bullock, J. S.,  et al. , 2001b, MNRAS, 321, 559B
\bibitem[Capelo et al.(2014)]{capelo+14} Capelo, P.~R., Volonteri, M., Dotti, M., et al.\ 2014, arXiv:1409.0004
\bibitem[Ceverino et al.(2012)]{ceverino+12} Ceverino, D., Dekel, A., Mandelker, N., et al.\ 2012, \mnras, 420, 3490
\bibitem[Choi et al.(2013)]{choi+13} Choi, J.-H., Shlosman, I., \& Begelman, M.~C.\ 2013, \apj, 774, 149
\bibitem[Choi et al.(2015)]{choi+15} Choi, J.-H., Shlosman, I., \& Begelman, M.~C.\ 2015, \apj, 450, 4411
\bibitem[Christodoulou et al. (1995)]{christodoulou+95} Christodoulou, D.M., Shlosman, I, \& Tohline, J.E. \ 1995, \apj, 443, 551
\bibitem[Cole et al. (1994)]{cole+94} Cole, S., et al.\ 1994, \mnras, 271, 718
\bibitem[Croton et al. (2006)]{croton+06} Croton, D.J., et al.\ 2006, \mnras, 365, 11
\bibitem[Davies et al.(2011)]{davies+11} Davies, M.~B., Miller, M.~C., \& Bellovary, J.~M.\ 2011, \apjl, 740, L42
\bibitem[A (2009)]{A2008} Dekel, A., \& Birnboim, Y., 2006, MNRAS, 368, 2
\bibitem[A (2009)]{A2008} Deng, H., Mayer, L., \& Meru, F., 2017, MNRAS, submitted
\bibitem[DevecchiVolonteri(2009)]{devecchi+09} Devecchi, B., \& Volonteri, M.\ 2009, \apj, 694, 302
\bibitem[Devecchi et al.(2012)]{devecchi+12} Devecchi, B., Volonteri, M., Rossi, E.~M., Colpi, M., \& Portegies Zwart, S.\ 2012, \mnras, 421, 1465
\bibitem[Debattista et al.(2006)]{debattista+06} Debattista, V.~P., Mayer, L., Carollo, C.~M., et al.\ 2006, \apj, 645, 209
\bibitem[De LuciaBlaizot (2007)]{delucia07} De Lucia, G., Blaizot, J., \ 2007, \mnras, 375, 2
\bibitem[A (2009)]{A2008} Di Cintio, A., et al. 2014, MNRAS, 437, 451
\bibitem[Di Matteo et al. (2005)]{dimatteo+05} Di Matteo, T., Springel, V., \& Hernquist, L., \ 2005, \nat, 433, 604
\bibitem[Di Matteo et al. (2012)]{dimatteo+12} Di Matteo, Khandai, N, de Graf, C. , Feng, Y., Croft, R.A.C., Lopez, J., \& Springel, V.,  \ 2012, \apj
\bibitem[A (2009)]{A2008} Di Matteo, T., Croft, R. A.C. , Feng, Y., Waters, D., \& Wilkins, S., 2017, MNRAS, 467, 4243
\bibitem[Dijkstra et al.(2006)]{dijkstra+06} Dijkstra, M., Haiman, Z., \& Spaans, M.\ 2006, \apj, 649, 14
\bibitem[Dijkstra et al.(2014)]{dijkstra+14} Dijkstra, M., Ferrara, A., \& Mesinger, A.\ 2014, \mnras, 442, 2036
\bibitem[Dotan et al.(2011)]{dotan+11} Dotan, C., Rossi, E.~M., \& Shaviv, N.~J.\ 2011, \mnras, 417, 3035
\bibitem[DurisenTohline(1985)]{durisen+85} Durisen, R.~H., \& Tohline, J.~E.\ 1985, Protostars and Planets II, 534
\bibitem[Durisen et al.(2007)]{durisen+07} Durisen, R.~H., Boss, A.~P., Mayer, L., et al.\ 2007, Protostars and Planets V, 607
\bibitem[Takeo et al. (2018)]{} Takeo, E., Inayoshi, K., Ohsuga, K., Takahashi, H, \& Mineshige, S.,  2018, \mnras, 476, 673
\bibitem[A (2009)]{A2008}Escala, A., 2007, ApJ, 671, 1264
\bibitem[Fan et al.(2001)]{fan+01} Fan, X., Narayanan, V.~K., Lupton, R.~H., et al.\ 2001, \aj, 122, 2833
\bibitem[A (2009)]{A2008} Feng, Y., Di Matteo, T., Croft, R. \& Khandai, N., 2014, MNRAS, 440, 1865
\bibitem[FerraraLoeb(2013)]{ferrara+13a} Ferrara, A., \& Loeb, A.\ 2013, \mnras, 431, 2826
\bibitem[A (2009)]{A2008} Ferrara, A., \& Latif, M., 2016, PASA, 33, e051
\bibitem[Fiacconi et al.(2013)]{fiacconi+13} Fiacconi, D., Mayer, L., Ro{\v s}kar, R., \& Colpi, M.\ 2013, \apjl, 777, L14
\bibitem[Fiacconi et al.(2014)]{fiacconi+14} Fiacconi, D., Feldmann, R., \& Mayer, L.\ 2014, arXiv:1410.6818
\bibitem[A (2009)]{A2008} Fiacconi, D. \& Rossi, E., 2016, MNRAS, 455, 2
\bibitem[A (2009)]{A2008} Fiacconi, D. \& Rossi, E., 2017, MNRAS, 464, 2259
\bibitem[A (2009)]{A2008} Fiacconi, D., Mayer, L,  Madau, P.. Lupi, A., Dotti, M. \& Haardt, F., 2017, MNRAS, 467, 4080
\bibitem[F{\"o}rster Schreiber et al.(2014)]{forster+14} F{\"o}rster Schreiber, N.~M., Genzel, R., Newman, S.~F., et al.\ 2014, \apj, 787, 38
\bibitem[Freitag et al.(2006)]{freitag+06} Freitag, M., G{\"u}rkan, M.~A., \& Rasio, F.~A.\ 2006, \mnras, 368, 141
\bibitem[A (2009)]{A2008} Governato, F., Brook, V, Mayer, L., et al., 2010, Nature, 463, 203
\bibitem[A (2009)]{A2008} Greif, T, Bromm, V., Clar, P.C. et al., 2012, MNRAS, 424, 39
\bibitem[A (2009)]{A2008} Johnnson, J. \& Haardt, F., 2017, PASA, 33, e007
\bibitem[Gammie(2001)]{gammie+01} Gammie, C.~F.\ 2001, \apj, 553, 174
\bibitem[A (2009)]{A2008}Georgiev, I. B\"oker, T, Leigh, N., L\"utzgendorf, N. \& Neumayer, N. 2016, MNRAS, 457, 2122
\bibitem[A (2009)]{A2008} Gillessen, S et al. 2017, ApJ, 837, 30
\bibitem[Girichidis et al. (2011)]{girichidis+11} Girichidis, P., Federrath, C., Banerjee, R. \& Klessen, R.S., \ 2011, \mnras, 413, 2741
\bibitem[Glebbeek et al.(2009)]{glebbeek+09} Glebbeek, E., Gaburov, E., de Mink, S.~E., Pols, O.~R., \& Portegies Zwart, S.~F.\ 2009, \aap, 497, 255
\bibitem[A (2009)]{A2008} Guedes, J. Madau, P., Mayer, L. \& Callegari, S., 2011, ApJ, 729, 125
\bibitem[Guo et al. (2011)]{guo+11} Guo, Q., et al., \ 2011, \mnras, 413, 101
\bibitem[A (2009)]{A2008} Greif, T.H., Glover, S.C.O., Bromm, V. \& Klessen, R.S., 2010, ApJ, 716, 510
\bibitem[G{\"u}rkan et al.(2004)]{gurkan+04} G{\"u}rkan, M.~A., Freitag, M., \& Rasio, F.~A.\ 2004, \apj, 604, 632
\bibitem[A (2009)]{A2008} Habouzit, M., Volonteri, M., Latif, M., Dubois, Y., \& Peirani, S., 2016, MNRAS, 463, 529
\bibitem[A (2009)]{A2008} Haemmerle, L., Woods, T. E., Klessen, R. S., Heger, A., Whalen, D, J., \  2018, MNRAS, 474, 2757
\bibitem[A (2009)]{A2008} Haiman, Z., 2013, ASSL, 396, 293
\bibitem[Hayfield et al.(2011)]{hayfield+11} Hayfield, T., Mayer, L., Wadsley, J., \& Boley, A.~C.\ 2011, \mnras, 417, 1839
\bibitem[He et al. (2018)]{} He, W., et al.\ 2018, PSAJ, 70, 33
\bibitem[Helled et al.(2013)]{helled+13} Helled, R., Bodenheimer, P., Podolak, M., et al.\ 2013, arXiv:1311.1142
\bibitem[Hernquist(1990)]{hernquist+90} Hernquist, L.\ 1990, \apj, 356, 359
\bibitem[Hernquist(1993)]{hernquist+93} Hernquist, L.\ 1993, \apjs, 86, 389
\bibitem[A (2009)]{A2008}Hirano, S., Hosokawa., T., Yoshida, N., Umeda, H., Omukai, K., Chiaki, G. \& Yorke, H. W. 2014, ApJ, 781, 60
\bibitem[Hoyle et al. (1963)]{hoyle+63} Hoyle, F. \& Fowler, W.A., \ 1963. \mnras, 125, 169
\bibitem[A (2009)]{A2008} Inayoshi, K., \& Haiman, Z., 2014, 445, 1549
\bibitem[A (2009)]{A2008} Inayoshi K., Haiman Z., Ostriker J. P., 2016, MNRAS, 459, 3738
\bibitem[Hopkins(2013)]{hopkins+13} Hopkins, P.~F.\ 2013, \mnras, 428, 2840
\bibitem[A (2009)]{A2008} Hopkins, P.F., 2015, MNRAS, 450, 53
\bibitem[Hosokawa et al.(2012)]{hosokawa+12} Hosokawa, T., Omukai, K., \& Yorke, H.~W.\ 2012, \apj, 756, 93
\bibitem[Hosokawa et al. (2013)]{hosokawa+13} Hosokawa, T., Yorke, H.W., Inayoshi. K., Omukai, K., \& Yoshida, N., \ 2013, \apj, 778, 178
\bibitem[A (2009)]{A2008} Jiang Y., F., Stone J. M., Davis S. W., 2014, ApJ, 796, 106
\bibitem[Jogee (2006)]{jogee06} Jogee, S., 2006, Lecture Notes in Physics,Berlin Springer Verlag, 693, 143
\bibitem[Johnson \& Bromm(2007)]{johnson+07} Johnson, J.L., \& Bromm, V., 2007, \mnras, 374, 1557
\bibitem[A (2009)]{A2008} Johnson, J.L., Whalen, D., Agarwal, B., Paardekooper, J.P. \& Kochfar, S., 2014, MNRAS, 445, 686
\bibitem[Kauffmann, White \& Guideroni (1993)]{kauffmann+93} Kauffmann., G., White, S.D.M., Guideroni, B., 1993, \mnras, 264, 201
\bibitem[Kaufmann et al.(2007)]{kaufmann+07} Kaufmann, T., Mayer, L., Wadsley, J., Stadel, J., \& Moore, B.\ 2007, \mnras, 375, 53
\bibitem[Kazantzidis et al.(2005)]{kazantzidis+05} Kazantzidis, S., Mayer, L., Colpi, M., et al.\ 2005, \apjl, 623, L67
\bibitem[Keller et al.(2014)]{keller+14} Keller, B.~W., Wadsley, J., Benincasa, S.~M., \& Couchman, H.~M.~P.\ 2014, \mnras, 442, 3013
\bibitem[Khochfar \& Burkert(2006)]{khochfar+06} Khochfar, S., \& Burkert, A.\ 2006, \aap, 445, 403
\bibitem[A (2009)]{A2008} Keres, D., Katz, N., Fardal, M., Dave, R., \& Weinberg, D., 2009, MNRAS, 395, 160
\bibitem[A (2009)]{A2008}Kormendy, J. \& Ho, L.C. 2013, ARA\&, 51, 511
\bibitem[A (2009)]{A2008} Kratter, C. \& Lodato, G, 2016, ARA\&A, 54, 271
\bibitem[Kritsuk et al.(2011)]{kritsuk+11} Kritsuk, A.~G., Norman, M.~L., \& Wagner, R.\ 2011, \apjl, 727, L20
\bibitem[Krumholz et al.(2006)]{krumholz+06} Krumholz, M.~R., Matzner, C.~D., \& McKee, C.~F.\ 2006, \apj, 653, 361
\bibitem[Latif et al.(2012)]{latif+12} Latif, M.~A., Schleicher, D.~R.~G., \& Spaans, M.\ 2012, \aap, 540, A101
\bibitem[Latif et al.(2013)]{latif+13} Latif, M.~A., Schleicher, D.~R.~G., Schmidt, W., \& Niemeyer, J.\ 2013, \mnras, 433, 1607
\bibitem[A (2009)]{A2008} Latif, M.A., \&  Schleicher, D.~R.~G., 2015, A \& A, 578, 118
\bibitem[A (2009)]{A2008} Latif, M.A. \& Volonteri, M., 2015, MNRAS, 452, 1026
\bibitem[A (2009)]{A2008} Latif, M. A., Omukai, K., Habouzit, M., Schleicher, D. R. G., Volonteri, M., 2016, 824, 40 
\bibitem[A (2009)]{A2008} Lin, D.N.C. \& Pringle, J.E., 1987, MNRAS, 225, 607
\bibitem[A (2009)]{A2008} Lodato, G., \& Rice, W.K.M., 2004, MNRAS, 351, 630
\bibitem[A (2009)]{A2008} Lodato, G., \& Rice, W.K.M., 2005, MNRAS, 358, 1489
\bibitem[Lodato \& Natarajan(2006)]{lodato+06} Lodato, G., \& Natarajan, P.\ 2006, \mnras, 371, 1813
\bibitem[Lupi et al.(2014)]{lupi+14} Lupi, A., Colpi, M., Devecchi, B., Galanti, G., \& Volonteri, M.\ 2014, \mnras, 442, 3616
\bibitem[A (2009)]{A2008} Lupi A., Haardt F., Dotti M., Fiacconi D., Mayer L., Madau P., 2016, MNRAS, 456, 2993
\bibitem[A (2009)]{A2008} Lynden-Bell, D., 1969, Nature, 223, 690
\bibitem[Madau \& Rees(2001)]{madau+01} Madau, P., \& Rees, M.~J.\ 2001, \apjl, 551, L27
\bibitem[Madau et al.(2014)]{madau+14} Madau, P., Haardt, F., \& Dotti, M.\ 2014, \apjl, 784, L38
\bibitem[Mayer \& Wadsley(2004)]{mayer+04} Mayer, L., \& Wadsley, J.\ 2004, \mnras, 347, 277
\bibitem[Mayer et al.(2005)]{mayer+05} Mayer, L., Wadsley, J., Quinn, T., \& Stadel, J.\ 2005, \mnras, 363, 641
\bibitem[Mayer et al. (2007)]{mayer+07} Mayer, L., Kazantzidis, S., Madau. P., Colpi, M., Quinn, T. \& Wadsley, J. \ 2007, Science, 316, 1874
\bibitem[Mayer et al.(2010)]{mayer+10} Mayer, L., Kazantzidis, S., Escala, A., \& Callegari, S.\ 2010, \nat, 466, 1082
\bibitem[Mayer(2013)]{mayer+13} Mayer, L.\ 2013, Classical and Quantum Gravity, 30, 244008
\bibitem[A (2009)]{A2008} Mazzucchelli, C., et al.\ 2017, \apj, 849, 91
\bibitem[Merloni \& Heinz (2008)]{merloni+08} Merloni, A., Heinz, S.,  2008, \mnras, 388, 1011
\bibitem[Meru \& Bate(2011a)]{meru+11a} Meru, F., \& Bate, M.~R.\ 2011, \mnras, 410, 559
\bibitem[Meru \& Bate(2012)]{meru+11b} Meru, F., \& Bate, M.~R.\ 2012, \mnras, 427, 2022
\bibitem[Milosavljevi{\'c} et al.(2009)]{milosavljevic+09} Milosavljevi{\'c}, M., Bromm, V., Couch, S.~M., \& Oh, S.~P.\ 2009, \apj, 698, 766
\bibitem[Mo et al.(1998)]{mo+98} Mo, H.~J., Mao, S., \& White, S.~D.~M.\ 1998, \mnras, 295, 319
\bibitem[Moody et al.(2014)]{moody+14} Moody, C.~E., Guo, Y., Mandelker, N., et al.\ 2014, \mnras, 444, 1389
\bibitem[Montero et al. (2012)]{montero+12} Montero, P., Janka, H.T. \& M\"uller, E.  \ 2012, \apj, 749, 37
\bibitem[A (2009)]{A2008} Moster, B. P., Naabm T, \& White, S.D.M., 2017, submitted to ApJ, arXiv:1705.05373
\bibitem[Mortlock et al. (2011)]{mortlock+11} Mortlock, D. J., et al., \ 2011, \nat, 464, 616
\bibitem[Navarro et al.(1996)]{navarro+96} Navarro, J.~F., Frenk, C.~S., \& White, S.~D.~M.\ 1996, \apj, 462, 563
\bibitem[Noguchi(1998)]{noguchi+98} Noguchi, M.\ 1998, \nat, 392, 253
\bibitem[A (2009)]{A2008}Omukai, K, ApJ, 546, 635
\bibitem[A (2009)]{A2008} Onorbe, J., et al., 2015, MNRAS, 454, 2092
\bibitem[A (2009)]{A2008} Pacucci, F,, Pallottini, A., Ferrara, A. \& Gallerani, S., 2017, MNRAS, 468, L77
\bibitem[Padoan et al. (1997)]{padoan+97} Padoan, P., Nordlund, A., \& Jones, B.J.T., \ 1997, \mnras, 288, 145
\bibitem[A (2009)]{A2008} Pardo, K., Goulding, A.D., Greene, J. E., et al. , 2016, ApJ, 831, 203
\bibitem[Pelupessy et al.(2007)]{pelupessy+07} Pelupessy, F.~I., Di Matteo, T., \& Ciardi, B.\ 2007, \apj, 665, 107
\bibitem[A (2009)]{A2008} Pezzulli, E, Valiante, R., Orofino, M. C, Schneider, R., Gallerani, S. \& Sbarrato, T., 2017a, MNRAS, 466, 2131
\bibitem[A (2009)]{A2008} Pezzulli, E. Volonteri, M., Schneider, R. \& Valiante, R., 2017b, MNRAS submitted,  2017arXiv170606592P
\bibitem[Pickett et al.(1996)]{pickett+96} Pickett, B.~K., Durisen, R.~H., \& Davis, G.~A.\ 1996, \apj, 458, 714
\bibitem[Portegies Zwart et al.(2004)]{portegies-zwart+04} Portegies Zwart, S.~F., Baumgardt, H., Hut, P., Makino, J., \& McMillan, S.~L.~W.\ 2004, Nature, 428, 724
\bibitem[A (2009)]{A2008} Press, W. H. \&  Schechter P., \ 1974, \apj, 187, 425
\bibitem[A (2009)]{A2008} Rafikov, R., 2005, ApJL, 621, L69
\bibitem[A (2009)]{A2008} Rafikov, R., 2013, ApJ, 774, 144
\bibitem[A (2009)]{A2008} Rafikov, R. 2015, ApJ, 804, 62
\bibitem[Regan \& Haehnelt(2009a)]{regan+09a} Regan, J.~A., \& Haehnelt, M.~G.\ 2009a, \mnras, 393, 858
\bibitem[Regan \& Haehnelt(2009b)]{regan+09b} Regan, J.~A., \& Haehnelt, M.~G.\ 2009b, \mnras, 396, 343
\bibitem[Reisswig et al.(2013)]{reisswig+13} Reisswig, C., Ott, C.~D., Abdikamalov, E., et al. \ 2013, Physical Review Letters, 111, 151101
\bibitem[Rice et al.(2003)]{rice+03} Rice, W.~K.~M., Armitage, P.~J., Bate, M.~R., \& Bonnell, I.~A.\ 2003, \mnras, 339, 1025
\bibitem[Ritchie \& Thomas(2001)]{ritchies+01} Ritchie, B.~W., \& Thomas, P.~A.\ 2001, \mnras, 323, 743
\bibitem[Ro{\v s}kar et al.(2014)]{roskar+14} Ro{\v s}kar, R., Mayer, L., Fiacconi, D., et al.\ 2014, arXiv:1406.4505
\bibitem[A (2009)]{A2008} Sadowski, A, Narayan, R., Tchekhovskoy, A., Abarca, D. , Zhu, Y \& McKinney, J.C., 2015, MNRAS, 447, 49
\bibitem[Saijo \& Hawke(2009)]{saijo+09} Saijo, M., \& Hawke, I.\ 2009, \prd, 80, 064001
\bibitem[Saitoh \& Makino(2013)]{saitoh+13} Saitoh, T.R. \& Makino, J. \ 2013, \apj, 768, 44
\bibitem[Saijo \& Hawke(2009)]{saijo+09} Saijo, M., \& Hawke, I.\ 2009, \prd, 80, 064001
\bibitem[Saitoh \& Makino(2013)]{saitoh+13} Saitoh, T.R. \& Makino, J. \ 2013, \apj, 768, 44
\bibitem[A (2009)]{A2008} Salpeter, E.E., 1964, ApJ, 140, 796
\bibitem[Scalo et al.(1998)]{scalo+98} Scalo, J., V{\'a}zquez-Semadeni, E., Chappell, D., \& Passot, T.\ 1998, \apj, 504, 835
\bibitem[A (2009)]{A2008} Seth, A., et al., 2008, ApJ, 687, 997
\bibitem[A (2009)]{A2008}Shakura, N.I, \& Sunyaev, R.A. 1973, A\&A, 24, 337
\bibitem[Shankar et al. (2004)]{shankar+04} Shankar, F., Salucci, P., Granato, G.L., DeZotti, G., Danese, L.  2004, \mnras, 354, 1020
\bibitem[A (2009)]{A2008} Shapiro, S., 2005, ApJ, 620, 59
\bibitem[A (2009)]{A2008} Shapiro, S., \& Teukolski.S.A., 1987, "Black holes, white dwarfs and neutron stars; The Physics of Compact Objects", ed . Wiley
\bibitem[Schleicher et al.(2013)]{schleicher+13} Schleicher, D.~R.~G., Palla, F., Ferrara, A., Galli, D., \& Latif, M.\ 2013, \aap, 558, A59
\bibitem[Shen et al.(2010)]{shen+10} Shen, S., Wadsley, J., \& Stinson, G.\ 2010, \mnras, 407, 1581
\bibitem[Shen et. al. (2007)]{shen07} Shen, Y., et al.  2007, \aj, I33, 2222
\bibitem[Shibata \& Shapiro(2002)]{shibata+02} Shibata, M., \& Shapiro, S.~L.\ 2002, \apjl, 572, L39
\bibitem[A (2009)]{A2008}Sobral, D., et al., 2015, ApJ, 808, 139
\bibitem[Somerville \& Primack (1999)]{somerville99} Somerville, R.S., Primack, J.R. \ 1999, \mnras, 310, 1087
\bibitem[Spaans \& Silk(2000)]{spaans+00} Spaans, M., \& Silk, J.\ 2000, \apj, 538, 115
\bibitem[Springel et al. (2005)]{springel+05} Springel, V., et al.  2005, \nat, 435, 629
\bibitem[Stahler \& Palla(2005)]{stahler+05} Stahler, S.~W., \& Palla, F.\ 2005, The Formation of Stars, by Steven W.~Stahler, Francesco  Palla, pp.~865.~I
\bibitem[Stinson et al.(2006)]{stinson+06} Stinson, G., Seth, A., Katz, N., et al.\ 2006, \mnras, 373, 1074
\bibitem[Tacconi et al. (2010)]{tacconi+10} Tacconi, L. J., Genzel, R., Neri, R., Cox, P., Cooper, M. C., Shapiro, K., Bolatto, A.,   Bouch\'e, N., Bournau
\bibitem[Tanaka \& Haiman(2009)]{tanaka+09} Tanaka, T., \& Haiman, Z.\ 2009, \apj, 696, 1798
\bibitem[Timlin et al. (2017)]{} Timlin, J. D., et al.\ 2017, arXiv:1712.03128
\bibitem[A (2009)]{A2008}Toomre, A., 1964. ApJ, 139, 1217
\bibitem[A (2009)]{A2008} Trakhtenbrot, B., et al. 2015, Science, 349, 168
\bibitem[Treister et al.(2013)]{treister+13} Treister, E., Schawinski, K., Volonteri, M., \& Natarajan, P.\ 2013, \apj, 778, 130
\bibitem[Van Wassenhove et al.(2014)]{vanwassenhove+14} Van Wassenhove, S., Capelo, P.~R., Volonteri, M., et al.\ 2014, \mnras, 439, 474
\bibitem[Vink(2008)]{vink+08} Vink, J.~S.\ 2008, New Astron. Rev., 52, 419
\bibitem[A (2009)]{A2008} Visbal, E., Haiman, Z., \& Bryan, G.,\ 2014, \mnras, 445, 1056
\bibitem[Volonteri et al.(2003)]{volonteri+03} Volonteri, M., Haardt, F., \& Madau, P.\ 2003, \apj, 582, 559
\bibitem[Volonteri \& Rees(2006)]{volonteri+06} Volonteri, M., \& Rees, M.~J.\ 2006, \apj, 650, 669
\bibitem[Volonteri \& Begelman(2010)]{volonteri+10} Volonteri, M., \& Begelman, M.~C.\ 2010, \mnras, 409, 1022
\bibitem[Walter et al. (2004)]{walter+04} Walter, F., et al., \ 2004, \apj, 615, L17
\bibitem[Wadsley et al.(2004)]{wadsley+04} Wadsley, J.~W., Stadel, J., \& Quinn, T.\ 2004, New A, 9, 137
\bibitem[A (2009)]{A2008} Weigel, A., Schawinski, K., Treister, E., Urrym C.M., Koss, M. \& Trakhtenbrot, B., 2015, MNRAS 448, 3167
\bibitem[A (2009)]{A2008} Wetzel, A., Hopkins, P. F., Kim, J.H.,, Faucher-Gigure, C.A., Keres, D., \& Quataert, E., 2016,ApJ, 827, L23
\bibitem[A (2009)]{A2008} White, S.D.M., \& Rees, M.J., 1978, MNRAS, 183, 341
\bibitem[A (2009)]{A2008} White, S.D.M., \& Frenk, C.S, 1991, MNRAS ,379, 52
\bibitem[A (2009)]{A2008} Wilkins, S. M., Feng, Y., Di Matteo, T., Croft, R. et al.\  2017, MNRAS, 469, 2517
\bibitem[Willott et al. (2010)]{willott+10} Willott, C., et al., \ 2010, AJ, 139, 906
\bibitem[Willott et al. (2010b)]{willott+10b} Willott, C., et al., \ 2010, AJ, 140, 546
\bibitem[Wise et al.(2008)]{wise+08} Wise, J.~H., Turk, M.~J., \& Abel, T.\ 2008, \apj, 682, 745
\bibitem[A (2009)]{A2008} Woods, T. E., et al., 2017, ApJL,  842, L6

\end{thebibliography}
\end{document}